\documentclass[iop]{emulateapj}
\usepackage{color}
\usepackage{amsmath}
\usepackage{mathrsfs}
\usepackage{stmaryrd}
\usepackage{multirow}
\usepackage{hyperref}
\usepackage{bm}

\def\be{\begin{equation}}
\def\ee{\end{equation}}
\def\bd{\begin{displaymath}}
\def\ed{\end{displaymath}}
\def\ba{\begin{aligned}}
\def\ea{\end{aligned}}

\def\bh{M_{\bullet}}

\def\g{{\rm g}}

\def\msun{M_{\odot}}
\def\AU{\rm ~AU}
\def\kms{\rm ~km~s^{-1}}

\def\DD{\bm{\Delta}}

\begin{document}

\title{Probing the spinning of the massive black hole in the Galactic Center via pulsar timing: 
A Full Relativistic Treatment}
\shortauthors{Zhang \& Saha}
\author{Fupeng Zhang$^{1,\dagger}$, Prasenjit Saha$^2$}
\affiliation{ $^1$ School of Physics and Astronomy, Sun Yat-Sen University, Guangzhou 510275, 
China;~$^\dagger$\, zhangfp7@mail.sysu.edu.cn; \\
$^2$ Physik-Institut, University of Z\"{u}rich, Switzerland, psaha@physik.uzh.ch}

\begin{abstract}
Pulsars around the Massive Black Hole (MBH) in the Galactic Center (GC) are expected to be 
revealed by the incoming  facilities (e.g., the Square Kilometre Array). 
Under a full relativistic framework with the pulsar approximated as a test 
particle, we investigate the constraints on the spinning of the MBH by monitoring the timing 
of surrounding pulsars. For GC pulsars orbiting closely around the MBH (e.g., $\la1000$AU), 
we find that full relativistic treatment in modeling accurately their timing signals can be necessary, 
as the relativistic signals are orders of magnitude larger than the time of arrival measurement 
accuracies. Although usually there are near-degeneracies among MBH spin parameters, 
the constraints on the spinning of the MBH are still very 
tight. By continuously monitoring a normal pulsar in orbits with a period of $\sim2.6$yr and 
an eccentricity of $0.3-0.9$ under timing precision of $1-5$ms, within $\sim 8$yr the spin 
magnitude and the orientations of the GC MBH can be constrained with $2\sigma$ error of $10^{-3}-10^{-2}$ 
and $10^{-1}-10^\circ$, respectively. Even for pulsars in orbits similar to the detected star S2/S0-2 
or S0-102, we find that the spinning of the MBH can still be constrained within $4-8$yr, with the most significant 
constraints provided near the pericenter passage. If the proper motion of the pulsars with astrometric accuracy of 
$10\mu$as can also be collected along with the timing measurement, then the position, velocity, mass 
and the distance to the Solar System of the MBH can be constrained about $\sim10\mu$as, 
$\sim1\mu$as$/$yr, $\sim 10\msun$ and $\sim1$pc, respectively. 
\end{abstract}

\section{introduction}

Due to their tremendous rotation stability, pulsars are believed to be
one of the best probes in testing gravity theories in various
astrophysical environments ~\citep[for reviews, see
  e.g.,][]{Stairs03,Will14,Lorimer08}.  Binary pulsars have provided
clean tests of various general relativistic (GR) effects, including
the decay of the orbital period by gravitational wave radiation
~\citep{Taylor94,Kramer06}.  Such systems present cases of
comparatively weak gravitational fields, i.e., with $GM/rc^2$ of
$10^{-5}$--$10^{-7}$, where $M$ is the mass of the system and $r$ is
the distance between the two components.  On the other hand, the
so-called ``S-stars'' near the Galactic Center (GC) \citep{Ghezetal08,
  Gillessenetal09,Gillessenetal17} delve into $\sim
GM/rc^2\simeq10^{-3}$ \citep[see, for example, Figure~1
  of][]{Angelil10a}.  In this region, $M\simeq 4\times10^6\msun$ is a
massive black hole (MBH).  If pulsars orbiting quite close to the MBH
could be found, the precise tracing of time of arrival (TOA) of their
pulses can be used for probing the Kerr spacetime near a black hole.

The existence of pulsars close to the GC MBH are inferred by the
discovery of hundreds of young and massive stars within the inner
parsec of GC~\citep[e.g.,][]{Paumard06,Lu13}. Some of those massive
stars (e.g., $\ga9\msun$) can leave neutron star remnants at the end of
their lifetime through supernova explosions.  The number of these
pulsars is expected to be about $100$ within the orbital period of
$100\,$yr~\citep[e.g.][]{Zhang14,Pfahl04,Chennamangalam14}.  The
innermost one of them could be in an orbit as tight as about
$\sim100$--$500\,\AU$ from the MBH~\citep{Zhang14}. The existence of a
population of normal pulsars in the GC has also been strongly suggested by
the magnetar recently revealed in this region, as magnetars are rare
pulsars~\citep[e.g.,][]{Rea13,Eatough13}.

The severe broadening of the pulse profile due to the hyper-strong
radio-wave scattering by the interstellar media in GC imposes
difficulties in revealing the existence of these objects. The search
of pulsars in GC has to be performed in high radio frequencies (e.g.,
usually $\ga 9$GHz)~\citep{Cordes97,Pfahl04}. Although a number of GC
pulsar searches have been performed, no normal pulsars have been
detected within the inner parsec so
far~\citep[e.g.,][]{Deneva09,Macquart10,Bates11}.  Future facilities,
e.g., Square Kilometre Array (SKA), may be able to reveal a number of
normal pulsars in this region, due to its very large collection area,
offering prospects of testing general relativity by their timing
observations~\citep[e.g.,][]{Shao15,Eatough15}.

Pulsar binaries with the GC-MBH would have some important difference
from known pulsar binaries.
\begin{itemize}
\item First, the orbital periods around the GC-MBH would be years or
  decades, compared to hours to weeks for stellar binaries with
  white dwarfs or neutron stars.  The same applies to S~stars.  As a
  result, orbital precession or indeed any orbit-averaged quantity is
  not the most useful variable.  The earlier literature on orbits
  around the GC-MBH (whether S~stars or pulsars) tended to focus on
  precession
  \citep[e.g.,][]{Jaroszynski98,RE01,Pfahl04,Will08,Merritt11,Liu12,Psaltis16}.
  More recent literature has emphasized relativistic effects that
  appear within a few orbits, especially near pericenter passages
  \citep{AS10,Angelil10a,Zhang15,Yu16}.
\item Second, in stellar-mass binaries it is essential to
consider the gravity of both bodies, which is typically done through
a post-Newtonian treatment~\citep{Blandford76,Damour86,Hobbs06}.
  For a pulsar-MBH binary, the mass ratio is much smaller
    ($\sim$ $10^{-7}-10^{-6}$) and the pulsar can be approximated by a
    test particle moving in a Kerr metric (see
    Section~\ref{subsec:effect_pulsar_mass} for the consequences of
    such an approximation).  
\item There are, however, Newtonian dynamical perturbations from all
  the other masses in the GC region~\citep{Merritt11,Zhang16}. 
  How to remove out this Newtonian ``foreground'' remains an unsolved 
  problem.  A possible filtering strategy using wavelets is suggested 
  by \citet{Angelil14}.
\end{itemize}

This work studies a pulsar in a Kerr metric.  The full relativistic
framework developed previously~\citep[][hereafter ZLY15]{Zhang15} to
simulate orbits and redshifts is modified to compute pulsar TOAs
instead. Note that although we describe the method as
``full relativistic treatment'' of the pulsar's motion, we have neglected 
the mass of the pulsar (For the difference if the pulsar's mass is not 
ignored see also Section~\ref{subsec:effect_pulsar_mass}.) 
and the orbital decay due to the gravitational wave radiation. 
By performing a large number of Markov Chain Monte Carlo (MCMC)
simulations, we investigate the constraints on the spinning of the MBH
for pulsars in various orbits and under different timing
accuracies.  Meanwhile, the proper motion of the pulsars
measured by radio astrometry could also be quite significant as they
are very close to the MBH. Here we investigate the possible benefits
of including the proper motion measurements of the pulsars, e.g., the
additional constraints on the mass, distance and proper motion of the
MBH.

This paper is organized as follows: In Section~\ref{sec:num_method} we
introduce the details of the numerical integration of the motion of
pulsar and the pulse trajectory from the pulsar to the observer under
the Kerr metric. By such a full relativistic method, we derive the
observables, i.e., the TOA and the proper motion of the target
pulsar. We have extended the previous relativistic
framework in ZLY15 by including the motion of the MBH itself in the
simulation. We investigate the spin-induced GR effects in pulsar
timing and proper motion for some hypothetical pulsars in
Section~\ref{sec:signals}. This section also compares with
orbit-averaged post-Newtonian theory for pulsar binaries,
and shows that such
approximations could deviate from the real evolution of the orbital
precession in the full GR case, with the differences that could be
apparent by timing measurements of pulsars in GC (see also
Section~\ref{subsubsec:example_pulsar}, or Figure~\ref{fig:fchq_spin_Ea}).
In Section~\ref{sec:spin_MCMC} we perform a large number of Markov
chain Monte Carlo simulations to investigate the constraints on
the parameters of the MBH, including the spin, mass, proper motion and
the GC distance. The discussion and the conclusions are shown in
Sections 5 and 6, respectively.

\section{Numerical methods}
\label{sec:num_method}
In this work, we adopt the full GR framework in ZLY15, which can
simulate both the motion of the particle around the massive black hole
and the propagation of the photons emitted from the particle (
star or pulsar) to the observer in pure Kerr metric.  The framework is
briefly described below, for the details we refer the reader to ZLY15.
We expand the framework to include the TOAs of pulsar in the
simulation, as well as the motion of the whole system with respect to
the solar system.  Details of these are given in
Section~\ref{subsec:sgra_motion} and Section~\ref{subsec:timing_pulsar}, 
respectively.

Alternative approaches would be a perturbative treatment similar to a
post-Newtonian expansion \citep[cf.,][]{Wex95} or a mixed perturbative
and numerical approach \citep{Angelil10a}.  The present method, though
complex in implementation, is conceptually simpler than these
approaches.
 
\subsection{The full Kerr metric framework}
\label{subsec:full_kerr_framework}
\begin{figure*}
\center
\includegraphics[scale=0.9]{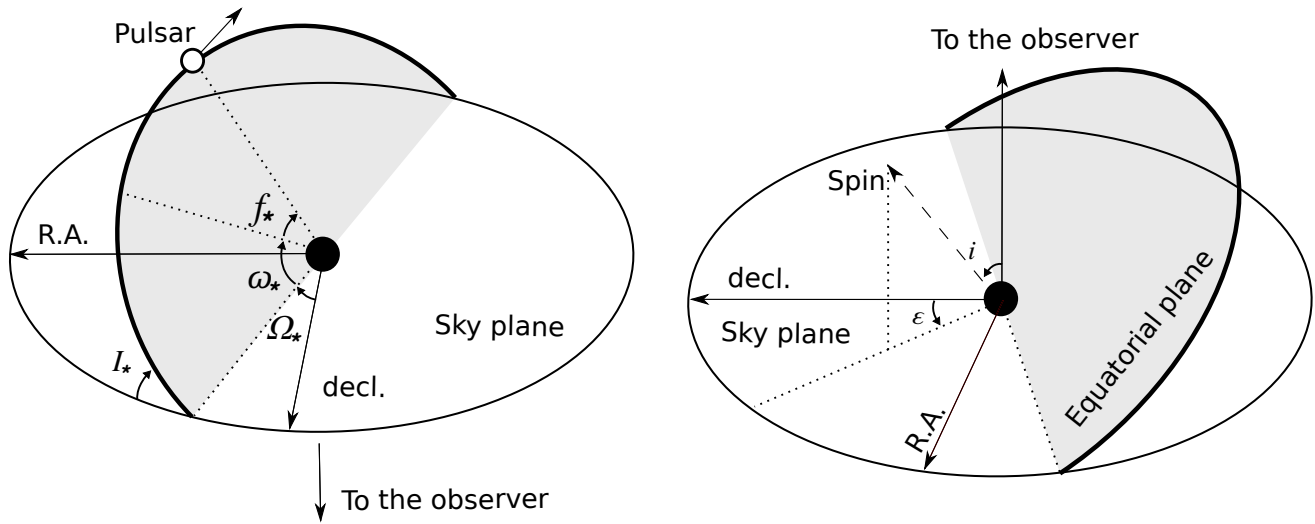}
\caption{Illustration of the angles used in this work. {\em Left
    panel:}~Angles of pulsar's orbit defined in the sky plane~\citep[see also][]
    {Eisenhauer05}. Here
  $I_\star$, $\Omega_\star$, $\omega_\star$ and $f_\star$ are the
  orbital inclination, position angle of the ascending node, angle of
  the periapasis, and true anomaly, respectively.  The orbital
  elements of the pulsar are defined with respect to the 
  position of the MBH, and here the MBH is located at $\rm R.A.=0$ and $\rm decl.=0$.
  {\em Right panel:}~Angles defined for the spin orientation: $i$ is
  the spin inclination with respect to the line of sight, while
  $\epsilon$ is the position angle of spin axis with reference to the
  direction of decl.  To make the angles easier to see, the two panels
  use different orientations --- note the direction to the observer in
  each case.}
\label{fig:angles}
\end{figure*}

We assume a Kerr spacetime around the GC, with a central mass of
$\bh=4\times10^6\msun$ corresponding to a gravitational radius of
$r_\g\simeq0.04\AU\simeq5\mu$as$\simeq2\times10^{-4}$\,mpc. The GC
distance we take to be $R_{\rm GC}=8$ kpc.

Orbits of pulsars are integrated in Boyer-Lindquist coordinates under
a full Kerr metric.  The equations of motion are given in Equations
19--22 in ZLY15. The orbital elements of the pulsar, being the
semimajor axis $a_\star$, eccentricity $e_\star$, inclination
$I_\star$, the position angle of the ascending node $\Omega_\star$,
angle of periapsis $\omega_\star$, and the true anomaly $f_\star$ (or
equivalently the time of the pericenter passage $t_{0\star}$), are all
defined with respect to the sky plane. The spin direction of the MBH
is defined by two angles: $i$ and $\epsilon$. Here $i$ is the
inclination to the line of sight, while $\epsilon$ is the position
angle with respect to a reference direction on the sky. All these
angles are illustrated in Figure~\ref{fig:angles}.

We adopt the backward light-tracing technique described in ZLY15 to
calculate the observed position $\mathbf{r}_\star$ of the target
pulsar in the sky plane (with respect to the position of the
MBH). Here $\mathbf{r}_\star=({\rm R.A.}, {\rm decl.})$, where ${\rm
  R.A.}$ is the right ascension and ${\rm decl.}$ is the declination. In
the simulations, ${\rm R.A.}=-\alpha r_{\rm g}/R_{\rm GC}$, ${\rm
  decl.}=\beta r_{\rm g}/R_{\rm GC}$, where $\alpha$ and $\beta$ are
dimensionless impact parameters of the pulse from the pulsar projected into 
the sky plane. These two values are given by
\be\ba
\lambda&=-\alpha\sin i, \\
q^2&=\beta^2+(\alpha^2-a^2)\cos^2 i.
\ea\ee
Here $\lambda=L_z/E_m$ and $q^2=Q/E_m^2$ are two constants of motion
for the pulse trajectory, $L_z$ is the azimuthal angular momentum, $Q$
is the Carter constant and $E_m$ is the energy of the photon at
infinity.

Similar to the case of stars, we can also obtain the corresponding
relativistic redshift $Z_\star$ of the pulse as
\be
1 + Z_\star =\frac{p_{\mu} U^\mu}{p_{\nu}^\text{obs} U^\nu_\text{obs}}
\ee
where $p^\mu$ is the four momentum of the pulse photon and $U^\mu$ is
the four velocity of the pulsar.  The redshift is not directly
measurable for pulsars, since the intrinsic frequency is not known,
but can be measured as it essentially is the derivative of the TOA (see
details in Section~\ref{subsubsec:similarity_toa_redshift}).

As we adopt the full Kerr metric, all the various GR effects on the
orbital motion of the target pulsar around the Kerr MBH and on the
propagation of the photons from the pulsar to the observer, are
simultaneously included in the mock observables of the target pulsar,
e.g., the TOA and the apparent motion in the sky plane.

\subsection{The motion of Sgr~A*}
\label{subsec:sgra_motion}
The position and the proper motion of the GC MBH in the sky plane is
usually indicated by its radio counterpart, i.e., Sgr~A*. The previous
framework of ZLY15 assumes that the GC MBH remains fixed in the sky.
Here we include the proper motion of the MBH in calculating the
apparent motion of the pulsars.  The apparent proper motion of a
pulsar in the sky plane is the sum of the motion of the MBH and also
its relative motion with respect to the MBH. By including 
the motion of the MBH in the modeling, the apparent motion of the pulsar 
can then be used directly in constraining the MBH parameters. The removal of 
the motion of Sgr~A* itself by some independent measurements is no longer
needed.  The relative acceleration between Sgr~A* and the Sun is
neglected.

By the method described in Section~\ref{subsec:full_kerr_framework}, 
we can obtain the evolution of the
relative sky position $\mathbf{r}_\star$ between the pulsar and the
MBH.  Suppose that Sgr~A* has a constant velocity in the sky plane
$\mathbf{V}_{\bullet}=(V_{x}, V_{y})$. The apparent sky position
of the Sgr~A*, $\mathbf{R}_\bullet$, is then given by
\be\ba
\mathbf{R}_\star&=\mathbf{r}_\star+\mathbf{R}_{\bullet,0}+\mathbf{V}_{\bullet}t_{\star}.
\ea\ee
Here $\mathbf{R}_{\bullet,0}=(X_{0}, Y_{0})$ is the initial position
of the Sgr~A* at the beginning of the observation, $t_\star$ is the
coordinate time of the pulsar. Note that the line of sight velocity
($V_z$) of the MBH respective to the Sun or the Local Standard of Rest
(LSR) can not be constrained in our model as it is absorbed in
the apparent pulse frequency. Here we simply assume that $V_z=0$.

The proper motion of Sgr~A* has been measured by a number of
observations~\citep[e.g.,][]{Reid04, Ghezetal08}.  The former give
values of $18\pm 7\kms$ in the galactic longitude and $-0.4\pm
0.9\kms$ in the galactic latitude (assuming $R_0=8$kpc).  If the
motions of the solar system and the LSR are included, the apparent
motion of the Sgr~A* is $-241\pm 15\kms$ and $-7.6\pm 0.7\kms$ in the
galactic longitude and latitude, respectively. In our MCMC
model-fitting procedure, in principle the position and the velocities
of Sgr~A* can be arbitrarily selected. However, to mimic the picture
expected in future observations, we adopt the values $V_x=-3.151$\,mas
yr$^{-1}$, $V_y=-5.547$\,mas yr$^{-1}$, both with respect to the Sun.
At the beginning of each simulation we set $X_0=0$\,mas and $Y_0=0$\,mas, 
but this makes no difference to the results. For simplicity, 
we ignore the apparent evolution of the geometric orientation of the orbital 
plane due to the proper motion of the barycenter of the pulsar-MBH binary
~\citep[See][]{Kopeikin96}.

\subsection{The timing of pulsars}
\label{subsec:timing_pulsar}

In the local frame of the pulsar, the proper time of the pulsar is
given by $\tau=\xi s$, where $s$ is the affine parameter (e.g.,
Equations 19-22 of ZLY15) and $\xi=m/E_m$, $m$ and $E_m$ being the rest
mass and the energy of the pulsar respectively.
In the local frame of pulsar, suppose that the pulse frequency is
$\nu_0$ at proper time $\tau_0$ and the first derivative of the pulse
with respect to proper time is given by $\dot \nu_0$, then in the
first order approximation, the phase of the emission $\phi$ for a
given proper time $\tau$ is
\be
\phi(\tau)=\nu_0 (\tau-\tau_0) + \frac{\dot \nu_0}{2} (\tau-\tau_0)^2.
\label{eq:phitau}
\ee
If we write $\nu=\nu_0+\dot \nu_0 (\tau-\tau_0)/2$ then we can also
express the above equation as $\phi(\tau)=\nu(\tau-\tau_0)$.  The zero
of proper time $\tau_0$ does not affect the results, so for simplicity
we set $\tau_0=0$. Note that when $\phi$ is exactly an integer, the
corresponding value of $\tau$ is the proper time of emission in the
local frame of pulsar.  Unless otherwise specified, we assume that
$\nu_0=2$~Hz (spinning frequency $0.5$\,s) and $\dot \nu_0=-10^{-15}$
s$^{-2}$ ($\sim 31.6$ nHz per year), which are typical values for a
normal pulsar.  The detection of millisecond pulsars is
unlikely~\citep[e.g.,][]{Cordes97,Macquart10}.

When a pulse is emitted at the proper time $\tau$, whenever $\phi$ in
Equation~\eqref{eq:phitau} is an integer, we can obtain the
corresponding coordinate time $t_\star$ of the pulsar from the
equations of motion (Equations 19--22 in ZLY15). According to the
light-tracing method described in ZLY15, we can trace the photon from
a distant observer (located at distance $r_0=10^8r_{\rm g}$) to the
position of the pulsar, and the time of propagation $t_{\rm prop}$ can
be obtained.  For simplicity we shift the time of propagation, i.e.,
$t_{\rm prop}\rightarrow t_{\rm prop}-r_0/c$, such that $t_{\rm prop}$
is approximately the additional time of propagation used for the pulse
crossing the pulsar-MBH binary respect to the position of the MBH. The
observed TOA $t_{\rm arr}$ of the pulse is then given by
\be
t_{\rm arr}=t_\star+t_{\rm prop}.
~\label{eq:tob}
\ee
We do not include the additional time corrections due to the
scatterings of the interstellar medium and the translations from the
barycenter of Solar system to the local time of observational stations
on earth. For more details of these corrections see~\citet{Edwards06}.

\section{The relativistic motion of Hypothetical Pulsars in the Galactic center}
\label{sec:signals}

With the numerical methods described in Section~\ref{sec:num_method}, here we investigate the GR signals, 
especially the spin-induced effects in the observables of hypothetical pulsars around the GC MBH. 
Let $\delta_a Y$ be the spin-induced difference on $Y$, here $Y$ is any quantity of interest, 
e.g., $t_{\rm arr}$, $\mathbf{R}_\star$, and etc. We can estimate $\delta_a Y$ by performing simulations 
with and without the spinning of the MBH and then estimate the difference in $Y$. More explicitly, 
$\delta_a Y$ is defined by
\be
\delta_a Y=Y(a,\bm{\vartheta})-Y(0,{\bm{\vartheta}})\simeq\frac{\partial Y}{\partial a} a
~\label{eq:deltaaY}
\ee
Here $\bm{\vartheta}$ stands for the initial values of all the parameters in the simulation except the MBH spin $a$.
For a given observational duration $T_{\rm tot}$, we define the spin-induced effects per orbit
as $\overline{\delta_a Y}$, which is given by
\be
\overline{\delta_a Y}=\frac{P}{T_{\rm tot}}
\left[\int_0^{T_{\rm tot}}\frac{1}{T_{\rm tot}} \left|\delta_a Y\right|^2
dt_{\rm arr}\right]^{1/2},
\label{eq:avgdaY}
\ee
Here $P$ is the orbital period. We estimate $\overline{\delta_a Y}$ by
performing simulations with duration of $T_{\rm tot}=3P$.  As roughly
$\delta_a Y\propto t_{\rm arr}/P$, the values defined above are
approximately independent on the observational duration.

If not otherwise specified, we always adopt $i=45^\circ$ and $\epsilon=180^\circ$ for the MBH spin 
orientation.


In the following section, we discuss plausible orbits of the pulsars in GC and assume 
the existence of some hypothetical pulsars within $\la 1000\AU$. 
The mock apparent position, TOA and the corresponding spin-induced effects for these 
example pulsars are described in Sections~\ref{subsec:appmotion_pulsar},~\ref{subsec:mock_pulsar_timing} 
and~\ref{subsec:spin_effect_pulsar}, respectively.

\subsection{The orbits of hypothetical pulsars}
\label{subsec:hypo_pulsars}
\begin{table}
\caption{Orbital parameters for example pulsars}
\centering
\begin{tabular}{lccccccccccc}\hline
\multirow{2}{0.8cm}{Name}    & \multicolumn{2}{c}{$a_\star$} &
\multirow{2}{0.5cm}{$e_\star$} & \multirow{2}{0.5cm}{$I_\star$} &
\multirow{2}{0.5cm}{$\Omega_\star$} & \multirow{2}{0.5cm}{$\omega_\star$} &
\multirow{2}{0.5cm}{$f_\star$} \\
\cline{2-3}
	&   AU\,$^a$ &  $r_{\rm g}\,^b$ &   & & & \\
\hline
S2-like  & 984 &  24949  & 0.88 & 135$\arcdeg$ & 225$\arcdeg$ &
63$\arcdeg$ & 180$\arcdeg$ \\
S0-102-like  & 848 & 21500  & 0.68 & 151$\arcdeg$ &
175$\arcdeg$ & 185$\arcdeg$ & 180$\arcdeg$ \\
Ea       & 300 & 7606  &  0.88 & 135$\arcdeg$ & 225$\arcdeg$ &
63$\arcdeg$ & 180$\arcdeg$ \\
%
Eb       & 300 &  7606 & 0.88 & 151$\arcdeg$ & 175$\arcdeg$ &
185$\arcdeg$ & 180$\arcdeg$\\
\hline
%
\end{tabular}
\tablecomments{ \,$^a$ in unit of AU.\\
\,$^b$ in unit of the gravitational radius $r_{\rm
g}=GM_{\bullet}/c^2$ and the MBH mass $M_{\bullet}$ is assumed to be
$4\times 10^6\msun$.\\
%
}
\label{tab:t1}
\end{table}
To explore the spinning the GC MBH the orbiting pulsar should be close
enough, e.g., at comparable distance to or closer than the currently
detected star S2/S0-2 or S0-102 ($a_\star\la 10^3$ AU). These pulsars
and their progenitors are unlikely to have formed in situ, as the
tidal forces of the MBH is quite strong.  Nor is it likely that they
migrated to these distance through secular dynamical process (e.g., by
two body relaxation processes) as the corresponding timescale is much
longer than the lifetime of the pulsar and the progenitor
stars. A plausible model is that their progenitors are
captured by the MBH through tidal break up of binary
stars~\citep{Zhang14}. By including the supernova kick,
secular dynamic relaxations and gravitational wave decay,
\citet{Zhang14} estimated that a number of $\sim10-15$ pulsars are
expected to be hidden within distance of $1000\AU$ from the black
hole.

\citet{Zhang14} show that the orbits of the hypothetical innermost pulsars are expected to have
$a_\star=100-500\AU$ and have distribution of eccentricity similar to those of the 
currently detected S-stars. Currently, the two closest S-stars from the GC MBH are the 
S2/S0-2~\citep{Gillessenetal09, Ghezetal08, Gillessenetal17} and the S0-102~\citep{Meyer12}. 
It is possible in some scenarios that the innermost pulsars have distances 
similar to these stars~\citep{Zhang14}, thus, it would be also quite interesting to see the 
spin-induced effects in timing of these pulsars, which are relatively away from the MBH and 
in periods of $11\sim15$yr. 

Based on these results, we select four example pulsars, of which 
the orbital parameters are shown in Table~\ref{tab:t1}. We notice that the GR and spin-induced effects depend 
on the orbital orientations, and thus two different orbital orientations of the example pulsars are 
assumed. We assume S2-like and S0-102-like pulsar, with the orbits similar to the currently detected 
S2/S0-2 and S0-102, respectively. Two miniature versions of these two pulsars, the pulsar ``Ea'' and ``Eb'' 
are with orbits similar to S2/S0-2 and S0-102, respectively, 
but both with $a_\star=300\AU$ (period of $\sim2.6$yr) and $e_\star=0.88$. 
Note that the orbital semimajor axis or the eccentricity of these 
example pulsars could vary according to the problems discussed in this work. Pulsars with orbits of 
$a_\star<100\AU$ and $e_\star>0.98$ may decay their orbits by gravitational wave within $10$Myrs
 (according to ~\citet{Peters64}). 
We avoid these pulsars as we have not included the effects of gravitational wave decay in the simulation. 

\subsection{The apparent motion and orbital precession of pulsar}
\label{subsec:appmotion_pulsar}
\begin{figure}
\center
\includegraphics[scale=0.55]{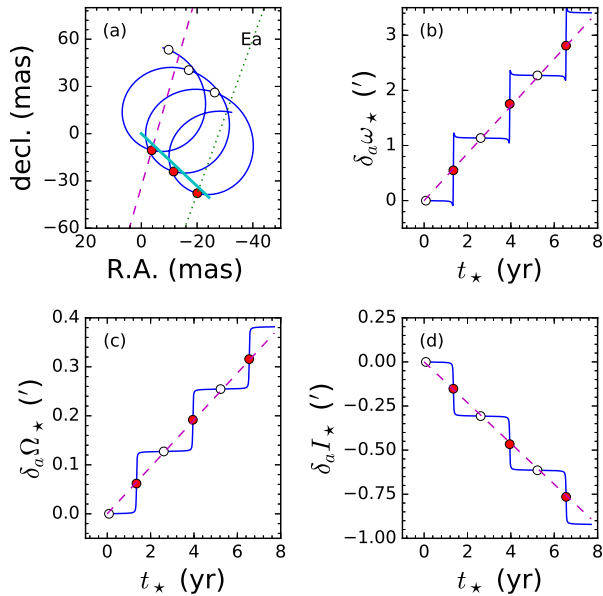}
\caption{Panel (a): Apparent motion of the example pulsar Ea in the
  the sky plane over three orbits ($\sim 8$yr). The magenta dashed and
  green dotted lines are the direction of the instantaneous
  eccentricity vector of the pulsar at the beginning and the end of
  the simulation, respectively. The cyan line shows the trajectory of
  the MBH in the GC.  Panels (b--d) The spin-induced orbital precession
  (in arcmin) in the angle of periapsis $\omega_\star$, the position angle of
  ascending node $\Omega_\star$ and the inclination $I_\star$.  The
  dashed magenta line show the theoretical expectations from
  Equation~\eqref{eq:orb_prec}.  The red solid and white empty circle in
  all panels show the position of the pericenter and apocenter passage
  points.  }
\label{fig:fimg}
\end{figure}

\begin{figure*}
\center
\includegraphics[scale=0.55]{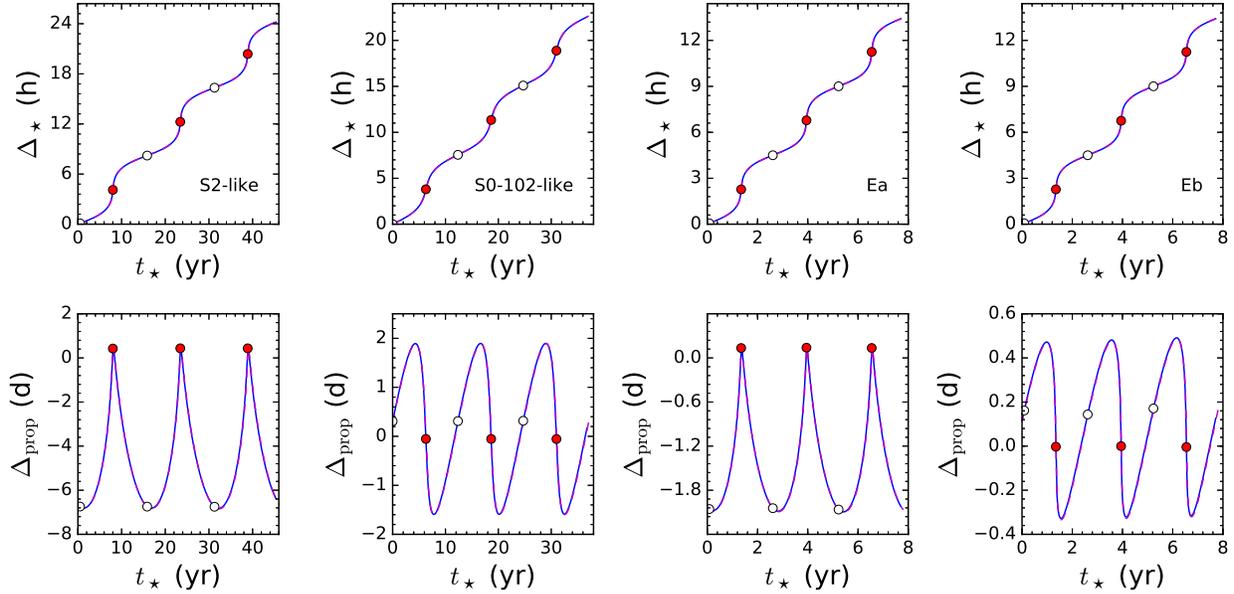}
\caption{Top panels: The Einstein delay (in hours) which connects the proper time to the 
coordinate time of the pulsar;
Bottom panels: The propagation time delay (in days) from a pulsar to the observer. 
The red solid and white empty circle show the position of the 
pericenter and apocenter passage points. The magenta dashed lines in the top panel shows the expectation 
from Equation~\eqref{eq:Edelay}, while the bottom panel from the sum of the Equation~\eqref{eq:Reodelay} 
and Equation~\eqref{eq:Shadelay}.
}
\label{fig:ftiming}
\end{figure*}

The top left panel of Figure~\ref{fig:fimg} show the apparent motion
in the sky plane of the example pulsar Ea in three orbits. The proper
motion of the Sgr~A* is quite significant. As the pulsar is moving
along with the Sgr~A*, its apparent trajectory in the sky plane shows a
spiral-like pattern. For the pulsar Ea, the transient velocity can be
up to $\sim100$mas/yr near the pericenter and $\sim10$mas/yr near the
apocenter, this suggests that the proper motion of a pulsar can be
easily measured by future long-baseline radio telescopes.

Both the TOA and the apparent positions of pulsars contain
relativity-induced effects. The Schwarzschild effects introduce 
precession only of the pericenter. Over one orbit, the precession is given by 
\citep{Wex99}
\be
\ba
\delta_S\omega&=6\pi G \bh c^{-2} a_\star^{-1}(1-e_\star^2)^{-1} 
\ea
\ee

The precession of orbits in a Kerr metric
is well known. If $\omega_\bullet,\Omega_\bullet,I_\bullet$ are
orbital elements with respect to the equatorial plane of a spinning
black hole, then over one orbit, the first two angles precess by
~\citep[e.g.,][]{Lense18,Wex99}
\be\ba
\delta_a \Omega_\bullet &=4\pi a (G \bh)^{3/2} c^{-3} a_\star^{-3/2}(1-e_\star^2)^{-3/2} \\
\delta_a \omega_\bullet &= - 3\delta_a\Omega_\bullet \cos I_\bullet
\label{eq:textbook-precession}
\ea\ee
to leading order. In observers' coordinates these precessions transform
\citep[cf.][]{Zhang15,Yu16}
\be\ba
\delta_a I_\star &= \sin i\cos(\epsilon-\Omega_\star)\,
                \delta_a\Omega_\bullet\,,\\
\delta_a \omega_\star &= \frac{\sin i \sin (\epsilon-\Omega_\star)}
                       {\sin I_\star} \delta_a\Omega_\bullet
                     + \delta_a \omega_\bullet\,, \\
\delta_a \Omega_\star &=
\left[\cos i -\frac{\cos I_\star \sin i \sin (\epsilon-\Omega_\star)}
      {\sin I_\star}\right]
\delta_a \Omega_{\bullet}\,.
\label{eq:dIOmom}
\ea\ee
For this, the direction cosine
\be
\cos I_\bullet = \cos I_\star \cos i+\sin I_\star \sin i
                 \sin (\epsilon-\Omega_\star)
\label{eq:spin-orient}
\ee
must be substituted into Equations \eqref{eq:textbook-precession}.
Combining the Schwarszchild and the spin effects, 
the orbital precession can be expressed as
\be\ba
\omega_\star&=\omega_{\star0}+\delta_a \omega_{\star}\frac{t_\star}{P}
+\delta_S\omega \frac{t_\star}{P},\\
\Omega_\star&=\Omega_{\star0}+\delta_a \Omega_{\star}\frac{t_\star}{P},\\
I_\star&=I_{\star0}+ \delta_a I_{\star} \frac{t_\star}{P}.
\label{eq:orb_prec}
\ea\ee
Here $t_\star\simeq \tau\simeq t_{\rm arr}$.

In fact, analytical expectation of Equation~\eqref{eq:orb_prec}
fails to trace the part of the orbital-element variations
that is caused by the spin-induced effects. 
Panels (b--d) of Figure~\ref{fig:fimg} show simulation results of
spin-induced orbital precession in $\Omega_\star$, $\omega_\star$ and
$I_\star$. The expectations from Equation~\eqref{eq:orb_prec} are shown in
the dashed magenta lines in each panel. Due to the relatively large
orbital eccentricity, the orbital precession of pulsar Ea in
simulations mainly occur near the pericenter, with nearly Keplerian
behavior near the apocenter.  In consequence, as we will show in more
detail later in Section~\ref{subsec:spin_effect_pulsar}, the
spin-induced TOA difference predicted by Equation~\eqref{eq:Bf76} below could
not trace accurately those obtained by our relativistic simulations. 
The deviations ($\sim 10$\,s) are quite apparent, considering the timing
accuracies ($\la1-10$\,ms) expected for future facilities, e.g., the
SKA.

\subsection{The timing of the pulsars}
~\label{subsec:mock_pulsar_timing}
According to Equation~\eqref{eq:tob}, for a proper time $\tau$ that corresponds to the emission of a pulse, 
its TOA can be alternatively expressed as 
\be
t_{\rm arr}=\tau+\Delta_\star+\Delta_{\rm prop}
\label{eq:tarr}
\ee
Here $\Delta_\star=t_\star-\tau$ and $\Delta_{\rm prop}=t_{\rm prop}$. $\Delta_\star$
translates the proper time to the coordinate time of pulsar. 
$\Delta_{\rm prop}=t_{\rm prop}$ translates the time of emission of 
the pulse in the local coordinate to the observer's frame. 

$\Delta_\star$ is also commonly dubbed the ``Einstein delay''. 
In the weak field approximation, $dt/d\tau\simeq 1+2G \bh/(rc^2)-G\bh/(2a_\star c^2)$.
By the method similar to~\citet{Blandford76}, the Einstein delay can be expressed as
\footnote{Note that the expression is slightly different
  from~\citet{Blandford76} and~\citet{Damour86} as we do not omit
  constants in the derivation.  In the literature, these constants are
  commonly absorbed in the spin frequency of pulsars.}  \be
\Delta_\star=t_\star-\tau\simeq \tilde{\gamma}(\sin E'-\sin
E'_0)+\frac{3 G \bh}{2a_\star c^2}t_\star
~\label{eq:Edelay}
\ee
Here $\tilde{\gamma}=\frac{G \bh P e_\star}{\pi a_\star c^2}$, 
$P=2\pi a_\star^{3/2} (G \bh)^{-1/2}$ is the orbital period of the pulsar,  
$E'$ is the eccentric anomaly which corresponds to a coordinate time $t_\star$, i.e., 
\be
E'-e_\star\sin E'=\frac{2\pi}{P}( t_{\star}-t_{0\star})
\ee
and $E'_0$ is the initial value of $E'$. 
From Equation~\eqref{eq:Edelay} we can 
see that $\Delta_\star$ is contributed by an oscillation term with the magnitude given by 
$\tilde \gamma$ and a linear term which is mainly due to the relativistic time dilation.

The top panels of Figure~\ref{fig:ftiming} show the evolution of $\Delta_\star$
obtained by the simulations (blue solid lines) and Equation~\eqref{eq:Edelay} (dashed magenta lines) 
for all example pulsars. We can see that the simulation results are well consistent with the analytical 
formula. $\Delta_\star$ appears to be the same for Ea and Eb as they have the same $a_\star$ and $e_\star$.
In three orbits, the delay mounts up to $\sim 13.5$ hour, and oscillates in magnitude of 
$\tilde \gamma\simeq 0.84$ hour. For the S2-like (or S0-102-like) pulsar, the delay 
mounts up to $\sim 24$ hour ($\sim 22$ hour) in $45$ years ($\sim 37$ years), and oscillates in magnitude of 
$\tilde \gamma \simeq 1.4$ hour ($\tilde \gamma \simeq 1$ hour). 

In weak fields, the last term in Equation~\eqref{eq:tarr}, $\Delta_{\rm prop}$, is approximately 
the sum of the ``Roemer delay'' $\Delta_{\rm R}$ and the ``Shapiro delay'' $\Delta_{\rm S}$,
i.e., $\Delta_{\rm prop}\simeq\Delta_{\rm R}+\Delta_{\rm S}$. 
The Roemer delay is the time used for a pulse propagating in a flat 
spacetime to the observer, which is given by~\citep{Damour86}
\be
\Delta_{\rm R}=\tilde{\alpha}(\cos E'-e_\star)+\tilde{\beta}\sin E'
~\label{eq:Reodelay}
\ee
Here 
\be
\ba
\tilde{\alpha}&=a_\star c^{-1}\sin I_\star\sin\omega_\star,\\
\tilde{\beta}&=(1-e_\star^2)^{1/2}a_\star c^{-1}\sin I_\star\cos\omega_\star,\\
\ea
\ee
The Shapiro delay is the additional time delay due to the curved spacetime, which is given by~
\citep{Shapiro64, Blandford76}
\be
\Delta_{\rm S}=\frac{2\bh G}{c^3}\ln\left[\frac{1+e_\star\cos f_\star}{1-\sin I_\star\sin (\omega_\star+f_\star)}\right]
~\label{eq:Shadelay}
\ee
For the GC MBH, we have $2\bh G/c^3\sim 39.3$\,s, which is also significant. 
Noticing that the orbit elements are precessing, thus 
in evaluating Equation~\eqref{eq:Reodelay} and~\eqref{eq:Shadelay} the orbital elements
$\omega_\star$ and $I_\star$ are replaced according to Equation~\eqref{eq:orb_prec}.

The bottom panels of Figure~\ref{fig:ftiming} show the evolution of  $\Delta_{\rm prop}$ 
of all the example pulsars. $\Delta_{\rm prop}$ oscillates as the pulsar rotates the MBH periodically, 
and is always dominated by the Roemer delay. For example pulsar Ea and Eb, the magnitude of 
the oscillation is $\sim2.3$ day and $\sim0.8$ day, respectively. For the S2-like pulsar and S0-102-like pulsar, 
the oscillation is $\sim 7.4$ day and $3.4$ day, respectively. The magenta lines in the left 
panels of Figure~\ref{fig:ftiming} show the model prediction from the sum of equation~\ref{eq:Reodelay} 
and~\ref{eq:Shadelay}. We can see that the simulations agree with the analytical formula well
\footnote{We find $\sim1\%$ difference between 
$\Delta_{\rm prop}$ (Similarly for the $\Delta_{\star}$) obtained 
by our numerical method and that obtained by the analytical method. The deviation
is mainly due to the difference of the metric adopted by these two methods (
the so-called ``gauge effect''). As a consequence, the positions, velocities and the timing of a pulsar with the same
initial conditions are slightly different. Note that the gauge effects
are automatically removed by comparing the spin-induced effects derived from the numerical 
and the analytical methods, thus they do not affect the results shown in top right panels of
Figure~\ref{fig:fchq_spin_Ea} and Figure~\ref{fig:fchq_spin_Eb}.}.

As a summary, the low order GR effects, e.g., the Einstein delay $\Delta_\star$ and Shapiro delay 
$\Delta_{\rm S}$, are both quite significant for all these example pulsars. 
Thus, it is expected that the low relativistic effects can be well tested if any pulsars within $\la 1000\AU$ from 
the MBH can be found. According to the measured Einstein and Shapiro delay, the MBH mass can also be well 
constrained. However, there is a near-degeneracy between the Einstein delay and the 
Roemer delay (See Equation~\eqref{eq:Bf76}), the Einstein delay could be separated 
only if significant changes of the orbit orientations are present.

\subsection{The spin-induced effects}
~\label{subsec:spin_effect_pulsar}
In this section, we discuss specifically the spin-induced signals on the observables of the pulsars, 
i.e., the TOA and the apparent motion in the sky plane. By numerical simulations and analytical arguments, 
we discuss the spin-induced effects for the example pulsars which is extracted according to 
Equations~\eqref{eq:deltaaY} and~\eqref{eq:avgdaY}. The details are shown in Section~\ref{subsubsec:example_pulsar}.
We also explore the spin-induced effects for pulsars in different orbital semimajor axes 
and eccentricities in 
Section~\ref{subsubsec:pulsar_df_orbits}. We notice that the spin-induced timing effects have some 
similarities with the corresponding redshift signals for stars. The details are shown in Section
\ref{subsubsec:similarity_toa_redshift}.
\subsubsection{Example pulsars}
\label{subsubsec:example_pulsar}
\begin{table}
\caption{The spin-induced effects of the example pulsars per orbit}
\centering
\begin{tabular}{lccccccccccc}\hline\\[-7pt]
Name & $\overline{\delta_a \Delta_\star}$ & $\overline{\delta_a \Delta_{\rm prop}}$ &
$\overline{\delta_a t_{\rm arr}}$ & $\overline{\delta_a \mathbf{R}_\star}$ \\
\hline
S2-like     & $0.067$\,s & $12$\,s    & $12$\,s       & $4.6\mu$as\\
S0-102-like & $0.024$\,s & $ 2.8$\,s  & $ 2.8$\,s     & $0.79\mu$as\\
Ea          & $0.22$\,s  & $22$\,s    & $22$\,s       & $8.6\mu$as\\
Eb          & $0.15$\,s  & $21$\,s    & $21$\,s       & $4.8\mu$as\\
\hline
%
\end{tabular}
%
%
\label{tab:effect_spin}
\end{table}

\begin{figure}
\center
\includegraphics[scale=0.55]{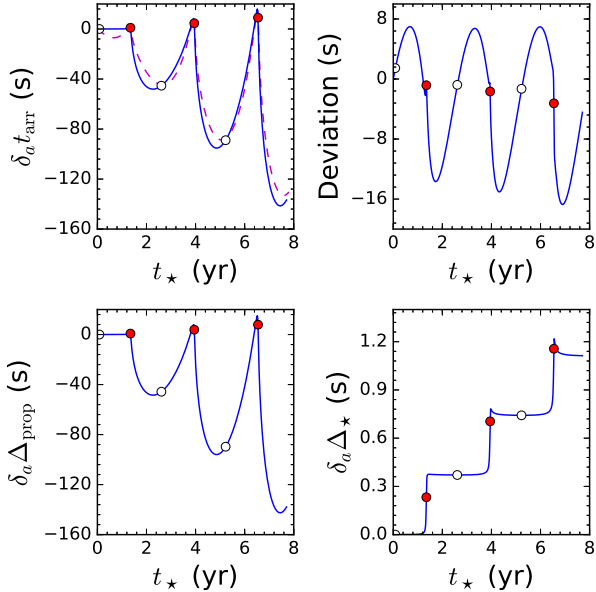}
\caption{Spin-induced relativistic effects on the TOA $\delta_\star t_{\rm arr}$ (top left panel),
the propagation time delay $\delta_a \Delta_{\rm prop}$ (bottom left panel) and 
the Einstein delay $\delta_a\Delta_{\star}$ (bottom right panel) of the example pulsar Ea. 
The blue solid line in top left panel shows the results of the 
numerical simulation while the dashed magenta lines show the 
theoretical expectations from Equation~\eqref{eq:Bf76}. Their differences as function of time 
are shown in top right panel. In all panels, the red solid and white empty circle show the position of 
the pericenter and apocenter passage points. }
\label{fig:fchq_spin_Ea}
\end{figure}

\begin{figure}
\center
\includegraphics[scale=0.55]{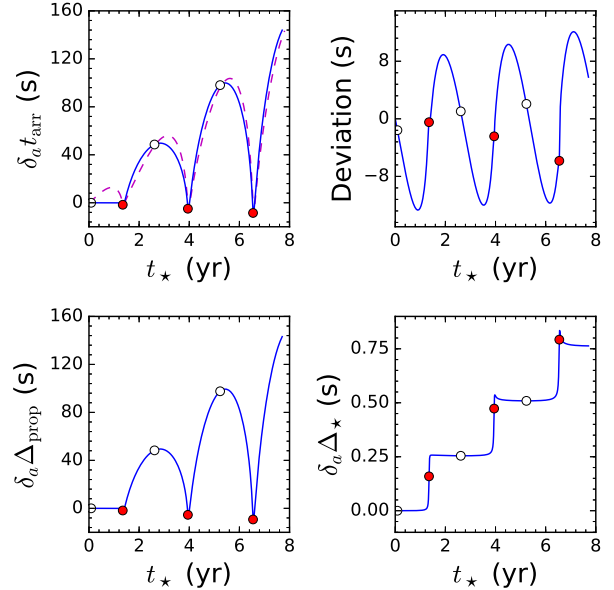}
\caption{Similar to Figure~\ref{fig:fchq_spin_Ea}, but for the example pulsar Eb.
}
\label{fig:fchq_spin_Eb}
\end{figure}

%
\begin{figure*}
\center
\includegraphics[scale=0.65]{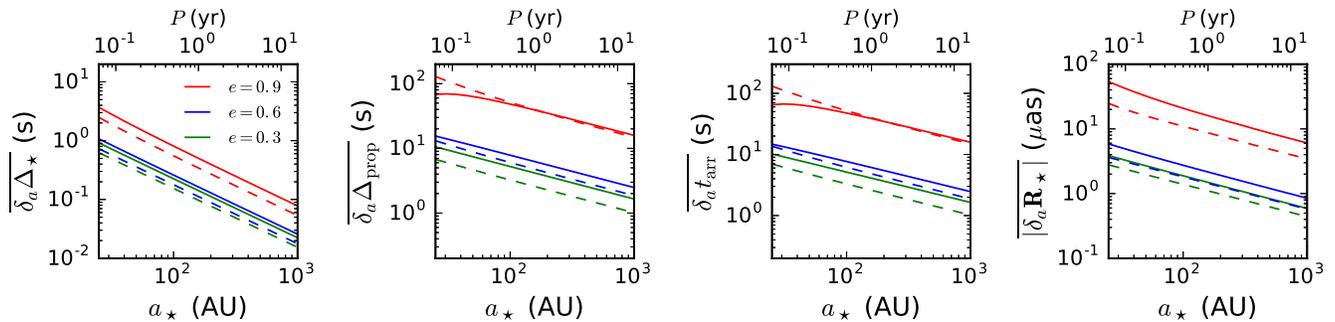}
\caption{Spin-induced difference per orbit on the Einstein delay $\overline{\delta_a \Delta_{\star}}$,
the propagation delay $\overline{\delta_a \Delta_{\rm prop}}$, TOA $\overline{\delta_a t_{\rm arr}}$
and the apparent position $\overline{|\delta_a \mathbf{R}_\star|}$ as function of the orbital semimajor axis.
The solid and dashed lines are results of the example pulsar Ea and Eb, respectively, in the cases of different
orbital eccentricity. 
}
\label{fig:fchq_t_R}
\end{figure*}

The TOA $t_{\rm arr}$ can be estimated by assuming weak fields and that the pulse propagates
in a straight line connecting the observer and the pulsar. The explicit relation between $t_{\rm arr}$
and $\tau$ in weak fields is given by~\citep{Blandford76,Taylor89}
\be\ba
t_{\rm arr}-\tau &\simeq\left[\tilde{\alpha}(\cos E-e_\star)+(\tilde{\beta}+\tilde{\gamma})\sin E
\right]\\
\times&\left[1+\frac{2\pi[\tilde{\alpha}\sin E-\tilde{\beta}\cos E]}
{P(1-e_\star\cos E)}\right]
~\label{eq:Bf76}
\ea\ee
Here $E$ is the eccentric anomaly, corresponding to a given TOA $t_{\rm arr}$ 
through
\be
E-e_\star\sin E=\frac{2\pi}{P}( t_{\rm arr}-t_{0\star})
\ee
Substituting the angles in Equation~\eqref{eq:orb_prec} to Equation~\eqref{eq:Bf76}, 
we could get the TOA when Schwarszchild and Lense-Thirling precession are included.
Note that if Equation~\eqref{eq:Bf76} is replaced with Equation 23~from \citet{Damour86},
the spin-induced effects are quite similar, as the low order effects, e.g., 
the Schwarszchild-induced effects, are subtracted according to Equation~\eqref{eq:deltaaY}.

According to Equation~\eqref{eq:tarr} we have $\delta t_{\rm arr}\simeq\delta_a\Delta_{\rm prop}+\delta_a\Delta_\star$.
Figure~\ref{fig:fchq_spin_Ea} shows the spin-induced difference in TOA $\delta_a t_{\rm arr}$, 
and the parts of propagation $\delta_a\Delta_{\rm prop}$ and of Einstein delay $\delta_a\Delta_\star$
for the example pulsar Ea. Similar results for Eb are shown in Figure~\ref{fig:fchq_spin_Eb}. 
In three orbits, the minimum and maximum TOA delay by spinning is 
$-141.5$\,s ($-10.9$\,s) and $15.99$\,s ($143.9$\,s) for Ea (for Eb), respectively. 
For S2-like (S0-102-like) pulsar, they are $-79.3$\,s and $8.47$\,s ($-5.21$\,s and $20.4$\,s), respectively. 
Such effects are maximal around the apocenter and minimum around the pericenter, with the rapid 
changes mainly caused by pericenter passages. The orbital-averaged values of them are shown in 
Table~\ref{tab:effect_spin}. We can see that $\delta_a\Delta_\star\ll\delta_a\Delta_{\rm prop}$, thus 
the spin-induced difference in the TOA is mainly contributed by the propagation part.
Notice that $\delta_a \Delta_{\rm prop}$ is approximately the sum of $\delta_a \Delta_{\rm R}$ 
and $\delta_a \Delta_{\rm S}$, which is the effect that the line of sight emitting position is 
changed by MBH spinning, and that the propagation trajectory is changed by MBH spinning, respectively. 

The spinning of the MBH changes the orbital 
motion of the pulsar compared to the case that the MBH is non-spinning, thus the pulsar feels a different 
potential and moves with different velocities, leading to difference in the Einstein delay.
The Einstein delay by spin-induced effects, i.e., $\delta_a\Delta_\star$, mounts up to a maximum value of
$1.2$\,s and $0.8$\,s in $8$ years for Ea and Eb, respectively. For S2-like and S0-102-like pulsar 
they are $0.37$\,s and $0.14$\,s respectively. The effects increase rapidly around the pericenter passages 
and remain almost constant near the apocenter. 
Note that both $\delta_a\Delta_{\rm prop}$ and $\delta_a\Delta_\star$ of
all example pulsars are quite significant compared to the typical timing accuracies of pulsar observations, 
e.g., $\la1-10$ms. This suggests that they can be measured quite accurately by future timing observations, if 
any pulsars with $a_\star\la1000\AU$ can be detected.

The model prediction of the spin-induced TOA difference can be obtained by first
replacing $\omega_\star$ and $I_\star$ in Equation~\eqref{eq:orb_prec}
into Equation~\eqref{eq:Bf76}, and then estimate the spin effects by Equation~\eqref{eq:deltaaY}. 
The magenta lines in top left panel of 
Figure~\ref{fig:fchq_spin_Ea} and~\ref{fig:fchq_spin_Eb} show the results for Ea and Eb, respectively. 
The analytical and simulation 
results are generally consistent, especially at the pericenter and apocenter passages. However, the discrepancies 
between the analytical and the simulation results are quite apparent in other regions. The maximum difference 
can be in orders of $\sim10$s (See the top right panel of Figure~\ref{fig:fchq_spin_Ea} and~\ref{fig:fchq_spin_Eb}), 
much larger than the measurement errors expected by the future telescopes. 
Such discrepancies arise as in analytical formula the orbital precessions are 
assumed to linearly increase with time. However, they does not describe accurately the evolution compared to those 
obtained by our relativistic simulations (See Figure~\ref{fig:fimg}). These results suggest that 
full relativistic treatment that is presented in this work, or more sophisticated analytical models which can 
trace accurately the orbital precession and describing the timing signals, are necessary of the GC pulsar timing observations.

\subsubsection{Pulsars in different orbits}
~\label{subsubsec:pulsar_df_orbits}
In this section, we explore the dependence of the spin-induced difference of observables on the distance and eccentricity 
of the pulsar. We perform simulations for pulsars similar to Ea or Eb, but their $a_\star$ varies
from $25-10^3\AU$ and $e_\star$ takes the values of $0.3$, $0.6$ or $0.9$. 
Figure~\ref{fig:fchq_t_R} shows the spin-induced effects per orbit according to Equation~\eqref{eq:avgdaY}.
We can see that these spin-induced timing differences show strong dependencies on the orbital distance.
For example, for the pulsars similar to Ea (or Eb) but with $e_\star=0.9$, 
$\overline{\delta_a \Delta_{\star}}$ varies from $3.6$\,s to $0.07$\,s,
(or $\overline{\delta_a \Delta_{\star}}$ varies from $2.4$\,s to $0.05$\,s)
if $a_\star$ changes from $25\AU$ to $10^3\AU$.
Similarly, $\overline{\delta_a t_{\rm arr}}$ varies from $65$\,s to $16$\,s 
(or $\overline{\delta_a t_{\rm arr}}$ varies from $130$\,s to $15$\,s) if $a_\star$ changes from $25\AU$ to $10^3\AU$.
Approximately, $\overline{\delta_a \Delta_\star} \propto a_\star^{-1}$, 
$\overline{\delta_a\Delta_{\rm prop}} \simeq\overline{\delta_a t_{\rm arr}} \propto a_\star^{-1/2}$.
The spin-induced timing differences depend also strongly on the orbital eccentricities. For example, 
$\overline{\delta_a t_{\rm arr}}$ and $\overline{\delta_a \Delta_{\star}}$  for pulsars with eccentricities of 
$e_\star=0.9$ are about one orders of magnitude larger than those with eccentricities of $e_\star= 0.6$ 
or $e_\star=0.3$. 

The right panel of Figure~\ref{fig:fchq_t_R} shows the spin-induced position difference per orbit, i.e.,
$\overline{|\delta_a\mathbf{R}_\star|}$, of pulsars at different orbital semimajor axes
and eccentricities. For pulsar Ea with eccentricity $0.9$ (or $0.3$), $\overline{|\delta_a\mathbf{R}_\star|}$ 
varies from $53\mu$as ($3.8\mu$as) to $17.8\mu$as ($0.6\mu$as) if $a_\star$ changes from $25\AU$ to $10^3\AU$. 
For pulsar Eb with $0.9$ (or $0.3$), $\overline{|\delta_a\mathbf{R}_\star|}$ from $25\mu$as ($2.8\mu$as) 
to $3.4\mu$as ($0.4\mu$as) when $a_\star$ varies from $25\AU$ to $10^3\AU$. Taking the astrometric accuracy achievable by future telescopes as $\sim 10\mu$as, we can see that the signal is only measurable
for pulsars with high orbital eccentricity (e.g., $e_\star\ga0.9$), or pulsars with orbital period less than one 
year, e.g., $a_\star\la100\AU$. For more details of the spin-induced position difference see~
\citet{Zhang15}.

Note that the orbital precession of pulsars with $a_\star\la100\AU$ and $e_\star=0.9$ is extremely strong: 
The orbital precession on $\omega_\star$ of these pulsar can be up to $\sim 10^\circ$ per orbit (which is the 
sum of the Schwarszchild and the Lense-Thirling precession). Such strong orbital precession causes
significant changes of the apparent configurations of the pulsar orbits. As consequences, 
$\overline{\delta_a t_{\rm arr}}$ and $\overline{|\delta_a\mathbf{R}_\star|}$ for these pulsars 
do not strictly follow the power law relation with $a_\star$ (See Figure~\ref{fig:fchq_t_R}).

\subsubsection{The TOA and the redshift signals}
\label{subsubsec:similarity_toa_redshift}
The correspondence between the timing and the redshift signal of the pulsar
can be obtained as follows. For two adjacent pulses separated in the local frame of the 
pulsar by $\delta \tau$, the TOA is separated in the observed frame by $\delta t_{\rm arr}$. 
When $\delta \tau\rightarrow0$ we have 
\be
\frac{dt_{\rm arr}}{d\tau}=\frac{Z_\star}{c}+1.
~\label{eq:dtobdtau}
\ee 
Therefore, the time of arrival can be estimated by integrating the above equation on the proper time $\tau$, i.e.,
$t_{\rm arr}\simeq \int (1+Z_{\star}/c) d\tau$. According to Equation~\eqref{eq:deltaaY}, 
the spin-induced TOA difference is then approximately given by
\be
\delta_a t_{\rm arr} \simeq \frac{1}{c}\int_0^{T_{\rm tot}}\delta_a Z_{\star}(\tau) d\tau.
~\label{eq:tarr_zshift}
\ee
The value of $\delta_a t_{\rm arr}$ is approximately given by 
the integration of the spin-induced redshift difference, i.e., $\delta_a Z_\star$, over time.
For pulsars comparably distant from the MBH as the S-stars, or closer in,
considering that $\delta_a Z\simeq1-10\kms\simeq10^{-5}-10^{-6}c$, 
we have $\delta_a t_{\rm arr}\simeq 10-10^2$s if 
$T_{\rm tot}$ is of order $\sim8$yr. These analysis results are consistent with our simulations.

\section{Constraints on the spin parameter from the motion of pulsars}
\label{sec:spin_MCMC}
In this section, we explore the achievable constraints on the spin of the GC-MBH 
from the TOA with and without supplementary observations of proper motions.
In Section~\ref{subsec:mcmc} we describe the 
details of the MCMC parameteric fitting method. The results for the constraints of the spin parameters and the other 
parameters of the MBH, e.g., proper motion, mass and distance, are shown in 
Section~\ref{subsec:MCMC_results_spin} and~\ref{subsec:MCMC_results_mr} respectively.
We also discuss the effects of the pulsar mass on our simulation results
in Section~\ref{subsec:effect_pulsar_mass}.
\subsection{The parameter-fitting method}
\label{subsec:mcmc}
We use the MCMC fitting method to study the constraints on the parameters of the pulsar-MBH binary. The initial conditions are provided by the following 17 parameters:
\begin{itemize}
\item Six parameters for the initial orbital elements of the pulsar: $a_\star$, $e_\star$, $I_\star$,  
$\Omega_\star$, $\omega_\star$ and $f_\star$.
\item  Two parameters describing the spin frequency of the pulsar: $\nu_0$ and $\dot \nu_0$. 
\item Four parameters for the black hole: mass $\bh$, spin magnitude $a$, spin inclination $i$, and spin position angle $\epsilon$.
\item Five parameters for the location and motion of the system with respect to the solar system: distance $R_{\rm GC}$, and proper motion $V_{\rm x},V_{\rm y}$ and
initial position $X_0,Y_0$ of the MBH on the sky plane. $V_{\rm z}$ does not appear as a separate parameter, as it is absorbed within $\nu_0$.
\end{itemize}
For a given set of mock observations, a $\chi^2$ value is computed and
a Bayesian posterior probability function is constructed in the usual
way, and then the Metropolis-Hasting algorithm is used to recover the
parameters with uncertainties. The prior on the spin parameters are all
flat distributions. The boundaries of them are $0<a<1$, $10<i<170^\circ$, 
$0^\circ<\epsilon<360^\circ$. Inclinations close to $0^\circ$ or $180^\circ$ are avoided as in these cases our
ray-tracing method can not calculate the trajectories of lights very 
accurately.
For the prior on other initial conditions we assume each of them 
a Gaussian prior with the scatter that is about three orders of magnitude larger than 
its converged width obtained by the MCMC simulation. The central value of the Gaussian distribution is its 
input value.

Note that for all the MCMC simulations shown in Section~\ref{subsec:MCMC_results_spin}
and Section~\ref{subsec:MCMC_results_mr} we set $a=0.6$ as the true value of the MBH spin. If $a=1$ is used instead, the constraints on the MBH-spin shown in Section~\ref{subsec:MCMC_results_spin} 
are about two times tighter, as the spin value is limited $0\le a\le1$.

To obtain meaningful constraints of these parameters through MCMC
fitting we need to use mock samples with several times more data
points than the number of free parameters. Thus, we use a total of
$N=120$ mock observations (we assume one TOA and astrometric data 
are collected per observation) for all the MCMC simulations shown in the
following sections, regardless of the observational duration assumed.
For example, if the observations last for three orbits, then mock
observables at 40 different epochs per orbit are collected. From
Figures~\ref{fig:fchq_spin_Ea} and~\ref{fig:fchq_spin_Eb} we can see
that each pericenter passage increases dramatically the magnitude of
the spin-induced effects in pulsar timing. Thus, to improve the
constraints on the spin of the MBH, the time intervals between each
observation is $\propto r^{-1.5}$, so that the orbit is more
frequently sampled near pericenter passages.

In the case that observations consist of TOA only, $\chi^2=\chi^2_{\rm
  T}$. Writing the $j$-th TOA as $t^{\rm arr}_{j}$, $\chi_{\rm T}^2$
is given by~\citep[cf.][]{Hobbs06}
\be
\chi^2_{\rm T}=\sum_{j=1}^{N}\frac{\{\phi[\tau(t^{\rm arr}_j)]-{\rm I}_j\}^2}
                                   {(\sigma_{j})^2} \,.
\label{eq:chqT}
\ee
Here 
\be
\sigma_{j}=\frac{d\phi}{dt^{\rm ob}}\sigma_T\simeq\nu \sigma_{\rm T}
\ee
is the measurement error in $\phi$, while $\sigma_{\rm T}$ is the TOA
measurement error.  The function $\phi[\tau(t^{\rm arr}_j)]$ is
provided by the full relativistic model and ${\rm I}_j$ is the integer
that is closest to $\phi[\tau(t^{\rm arr}_j)]$.

The expected timing accuracy of pulsar for SKA of $\sim1$ hour
integration can be down to $\sigma_T\simeq100\mu$s~\citep{Liu12} if
the frequency can be up to $\ga15\,$GHz, and $\sigma_T\simeq0.1-10\,$ms
if the frequency is between $\ga 5\,$GHz and $\la 15\,$GHz.
Considering that many factors can limit the measurement accuracy, we
could expect that the TOA measurement accuracy may vary between
$0.5-50\,$ms.  This level is assumed in our simulations.  The
numerical accuracy in the simulated TOA is much higher ($\la
10\,\mu$s) so as to avoid any contamination due to numerical
errors.  If $\sigma_T\la0.5\,$ms is assumed, the constraints on the
spin and other parameters of the MBH are correspondingly tighter.

Astrometric measurements of the pulsar, if they are available, can
also be used in the MCMC runs along with the timing measurements. Suppose
that each observation is denoted by $\mathbf{R}^{\rm ob}_{\star,j}$,
for $j=1,\cdots, N$. The chi-square value can be expressed as
$\chi^2=\chi^2_{\rm T}+\chi^2_{\rm P}$. Here
\be
\chi^2_{\rm P}=\sum_{j=1}^{N}
\frac{(\mathbf{R_\star}-\mathbf{R}^{\rm ob}_{\star,j})^2}{\sigma^2_{\rm p}},
\label{eq:chqR}
\ee
where $\sigma_{\rm p}$ is the astrometric error. Note that for both the mock timing and the astrometric data
we assume that the measurement errors are all Gaussian. 

The SKA is expected to operate with the baselines up to $3000\,$km,
and thus its image resolution could be up to $2\,$mas at
$10\,$GHz~\citep{Godfrey12}. The astrometric accuracy could be even
higher, of order $\sim10\,\mu$as~\citep{Fomalont04}. In this work, we
simply assume that $\sigma_{\rm P}=10\,\mu$as.  If lower astrometric
accuracies are assumed, the constraints on the spin of the MBH are
slightly affected for those pulsars with $a_\star\la 300\,\AU$, as the
contribution of the astrometric measurements for these pulsars to the
fitting are much smaller than those of the timing, i.e., $\chi^2_{\rm
  T}\gg \chi^2_{\rm P}$. For other parameters of the MBH, e.g., $\bh$,
$R_{\rm GC}$ or the proper motions, the constraints on them will be
correspondingly weaker if one sets $\sigma_{\rm P}>10\,\mu$as.

When only the TOAs of pulsars are used, there is no information on
$\Omega_\star$ and $R_{\rm GC}$, hence these must be excluded from the
fitting procedure.  The orbital inclination $I_\star$ is classically
degenerate with $a_\star$.  The Einstein delay breaks this degeneracy
\citep[cf.][]{AS11} but as the effect is small, and a strong
correlation remains, which significantly slows the convergence of the
MCMC procedure.  In order to concentrate on the spin, we fixed
$I_\star$ in the simulations with TOA only, leaving 10 parameters to
fit.

In the cases that both the TOA and the proper motion of pulsars are
included in the MCMC fitting, we can set all the 17 parameters in the
system to be free. In this case, we can constrain the proper motion of the
MBH. We find that the constraints on the spin parameters depend
somewhat on whether the proper motion of the MBH is taken as free or
not. Thus, for some of the MCMC simulations we fix the position and proper 
motion of the MBH, leaving only $13$ free parameters. 

\subsection{Constraints on the spin parameters of the MBH}
~\label{subsec:MCMC_results_spin}

\begin{figure}
\center
\includegraphics[scale=0.5]{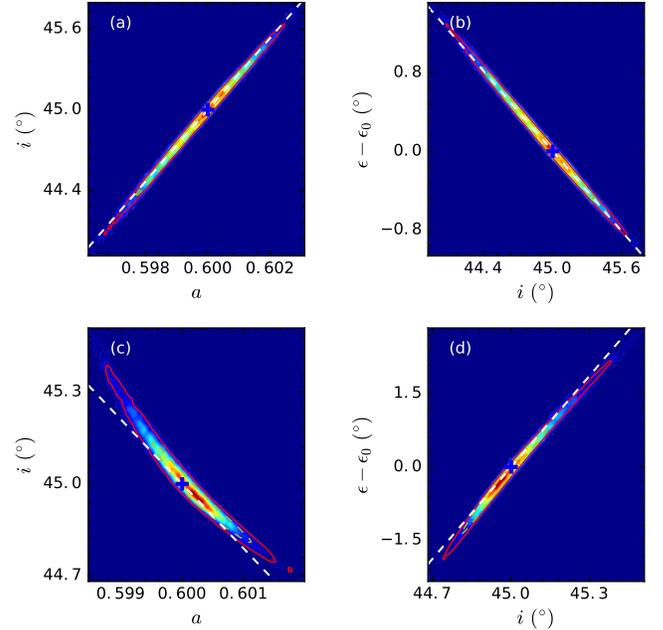}
\caption{Constraints on the MBH spin parameters when only the TOAs of
  the pulsar are fitted.  The correct values are marked by blue
  crosses.  The assumed timing accuracy is $\sigma_{\rm
    T}=5\,$ms. Panels~(a) and (b) show results for the example pulsar
  Ea, while Panels~(c) and (d) show results for Eb. The dashed white
  lines show parameter degeneracies at leading order: from
  Equation~\eqref{eq:da_di} in the left panel and from
  Equation~\eqref{eq:de_di} in the right panel.  The color contour maps
  represent the mean likelihood of the MCMC sample, and the line
  contours represent the marginalized distribution. The yellow dashed
  and red solid lines in each panel show the $1\sigma$ and $2\sigma$
  confidence levels respectively.}
\label{fig:mcmc_pure_pulse}
\end{figure}
By the MCMC methods described above, we perform a large number of MCMC
runs to investigate the constraints on the spinning and other properties 
of the MBH by monitoring pulsars shown in Table~\ref{tab:t1}. 
We find that near degeneracies appear among spin parameters, however, 
the constraints of spin can still be very tight. 
Even for S2-like or S0-102-like pulsars, the spin of the MBH (whether 
the MBH is spinning or not) can still be probed within $\sim4-8\,$yr and 
$\sim2-4\,$yr in optimistic scenarios. The details of the 
degeneracies and the constraints on spin parameters can be found in 
Section~\ref{subsubsec:degeneracy} and~\ref{subsubsec:example}, 
respectively. 

We expand these studies for pulsars to other conditions of orbits and
measurement accuracies. By performing a large number of MCMC
simulations we discuss how the constraints on spin parameters of the
MBH change for pulsars with different semimajor axes $a_\star$, 
eccentricities $e_\star$, and other parameters. 
For more details see Section~\ref{subsubsec:general}.

\subsubsection{Near-degeneracies among spin parameters}
\label{subsubsec:degeneracy}
\begin{figure*}
\center
\includegraphics[scale=0.6]{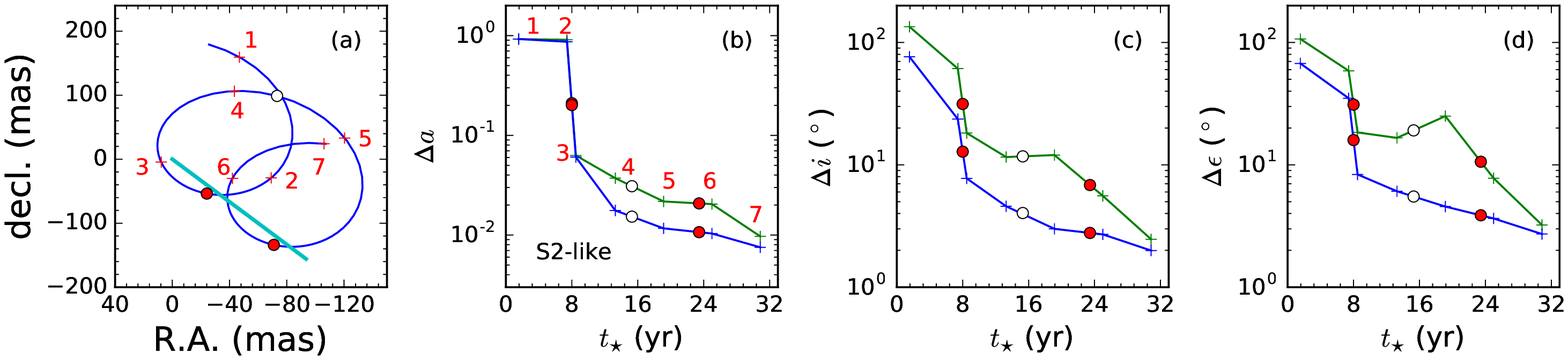}
\caption{Constraints on the MBH spin parameters from an S2-like
  pulsar, as a function of the observational duration. Panel~(a) show
  the apparent trajectories on the sky: blue curve for the pulsar and
  cyan line for the MBH.  Initial conditions are given in
  Table~\ref{tab:t1}, and the assumed measurement errors are
  $\sigma_{\rm T}=5\,$ms and $\sigma_{\rm p}=10\mu\,$as. We tried
  seven different observational durations, and the corresponding
  position of the pulsar at the end of observation are marked by
  numbers in Panel~(a).  Each of these seven cases has 120 mock
  observations.  The accuracy on the recovered parameters are shown in
  Panel (b)--(d): green lines show results when only TOAs are used,
  while blue lines show results when both the timing and the proper
  motion are used. The red filled and empty white circles mark the
  position where the pulsar passes pericenter and apocenter
  respectively.}
\label{fig:mcmc_fits_S2_like}
\end{figure*}
\begin{figure*}
\center
\includegraphics[scale=0.6]{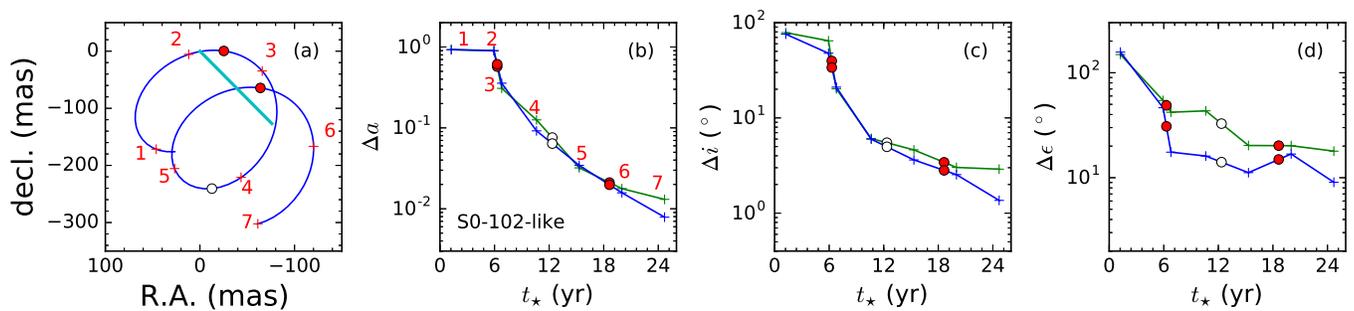}
\caption{Similar to Figure~\ref{fig:mcmc_fits_S2_like}, but for an
  S0-102-like pulsar.  }
\label{fig:mcmc_fits_S0-102_like}
\end{figure*}

Figure~\ref{fig:mcmc_pure_pulse} shows the constraints on the spin
magnitude and orientation of the MBH from observing the hypothetical
example pulsars Ea and Eb with $\sigma_{\rm T}=5\,$ms.  The
constraints on $a$, $i$ and $\epsilon$ are nearly degenerate. 
The near degeneracies among the spin parameters when only the timing 
signals are used can be understood as follows.  Let us write
\be\ba
\epsilon'&\equiv\epsilon-\Omega_\star,\\
a_x&=a\sin i\cos \epsilon', \\
a_y&=a\sin i \sin \epsilon',\\
a_z&=a\cos i
\label{eq:axayaz}
\ea\ee

The observed TOAs of a pulsar contain information on spin-induced
precessions $\Delta I_\star$ and $\Delta \omega_\star$ but not on
$\Delta \Omega_\star$ in Equation~\eqref{eq:dIOmom}.  Using that fact,
and substituting from Equations~\eqref{eq:textbook-precession} and
\eqref{eq:spin-orient} we have
\be
\ba
a_x &= {\rm const,}\\
a_z-\Gamma a_y &= {\rm const.}
\ea
\label{eq:degen-axayaz}
\ee
Here
\be
\Gamma=\frac{1-3\sin^2 I_\star}{3\cos I_\star \sin I_\star}.
\label{eq:degen-gamma}
\ee
Now using the notation $\bm{\delta}Y$ for fitting uncertainty in any
quantity $Y$, Equation~\eqref{eq:degen-axayaz} gives two relations
involving $\bm\delta a_x$, $\bm\delta a_y$ and $\bm\delta a_z$.  With
the help of Equation~\eqref{eq:axayaz} these can be rearranged as
two relations
\be
(\Gamma \sin i-\cos i \sin \epsilon')\bm{\delta} a=
-(\Gamma \cos i+\sin i\sin \epsilon')a \bm{\delta} i
\label{eq:da_di}
\ee
and
\be
\cos \epsilon' \bm{\delta} i=-(\Gamma \sin i
-\cos i\sin \epsilon')\sin i \bm{\delta} \epsilon \,.
\label{eq:de_di}
\ee
Note that if $a\simeq1$, we have the restriction that $\bm{\delta}a\le1-a$.

The white dashed lines in Figure~\ref{fig:mcmc_pure_pulse} show the
predictions from Equation~\eqref{eq:da_di} and~\eqref{eq:de_di}.
 They are in good agreement with the near-degeneracies in the simulation
results.  Discrepancies appear as Equations~\eqref{eq:da_di}
and~\eqref{eq:de_di} are first order approximations.

From Equations~\eqref{eq:dIOmom} and \eqref{eq:degen-gamma} we can see
that the degeneracies can be broken in the following four cases:
(1)~$\delta_a \Omega_\star$ can be inferred if proper motion of the pulsar can
be measured, thus the degeneracies are broken if it 
is measured with considerable accuracy such that $\chi_{\rm P}\sim\chi_{\rm T}$. 
For an S2-like or S2-102-like pulsar the degeneracies are slightly 
weakened by additionally including the astrometric measurements. However, 
for pulsars Ea and Eb, similar degeneracies appear as the constraints are still
dominated by the parts from timing, i.e., $\chi_{\rm P}\ll\chi_{\rm T}$. 
(2)~According to Equation~\eqref{eq:degen-gamma}, the degeneracies are
only functions of the orbital inclination $I_\star$ of the
pulsar. Thus, the degeneracy can be broken if the apparent orbital precession
of $I_\star$ is significant; (3) Similarly, degeneracies can be broken
by combining the timing of another pulsars with different inclination.
(4)~The quadrupole-moment effects are strong enough that they provide
another independent constraint on spin.  We can see that (2) and (4)
can be naturally satisfied if a pulsar has a short orbital period
(e.g. $\la0.5\,$yr), or the duration of the timing observation is long
enough.

\subsubsection{Results from the example pulsars}
\label{subsubsec:example}
\begin{table}
\caption{Constraints on the spin parameters from example pulsars}
\centering
\begin{tabular}{lccccccccccc}\hline
\multirow{2}{1.0cm}{Name} & \multicolumn{3}{c}{TOA$^c$} &
& \multicolumn{3}{c}{TOA $+$ astrometric$^d$} \\
\cline{2-4} \cline{6-8}
	&  $\DD a$ & $\DD i$  & $\DD\epsilon$ & 
	&  $\DD a$ & $\DD i$  & $\DD\epsilon$ \\
\hline
S2-like$^a$  & $0.0098$ & $2.5^\circ$ & $3.2^\circ$      & & $0.0075$ & $2.0^\circ$ & $2.7^\circ$\\
S0-102-like$^a$  & $0.0130$ & $2.9^\circ$ & $17.9^\circ$ & & $0.0079$ & $1.4^\circ$ & $9.1^\circ$ \\
Ea$^b$       & $0.0046$ & $1.2^\circ$ & $1.7^\circ$      & & $0.0044$ & $1.2^\circ$ & $1.7^\circ$\\
Eb$^b$       & $0.0023$ & $0.5^\circ$ & $3.3^\circ$      & & $0.0033$ & $0.7^\circ$ & $4.7^\circ$\\
\hline
%
\end{tabular}
\tablecomments{\,$^a$ Constraints from an S2-like or S0-102-like pulsar in two orbits, corresponding to $\sim30\,$yr or $\sim24\,$yr, respectively.\\
\,$^b$ Constraints from Ea or Eb in three orbits, corresponding to $\sim8\,$yr.\\
\,$^c$ Using only the timing of pulsars.\\
\,$^d$ Using both the timing and the apparent motion of pulsars.\\
}
\label{tab:constrain_spin}
\end{table}
Although near degeneracies appear among spin parameters, 
pulsars can still deliver very tight constraints on the
spin parameters, even though the timing accuracy, i.e., $\sigma_{\rm
  T}=5$ms, is not the most optimistic value. To quantify the
constraints on any quantity of interest $Y$, let $Y_{2+}$ and $Y_{2-}$
be the upper and lower $2\sigma$ confidence limits.
Then let
\be \DD Y=Y_{2+}-Y_{2-}
\ee
will be the $2\sigma$ range of $Y$. The constraints $\DD Y$ on the
spin of the MBH from all the example pulsars are given in
Table~\ref{tab:constrain_spin}. We can see that $\DD a$, $\DD i$ and
$\DD \epsilon$ are of order $10^{-3}-10^{-2}$, $0.5-3^\circ$,
$1-20^\circ$, respectively. 

For an S2-like or S0-102-like pulsar, we find that the constraints of
spin from observations over two orbital periods, i.e., $\sim 30\,$yr
or $\sim 25\,$yr, are also tight. Figure~\ref{fig:mcmc_fits_S2_like} and
Figure~\ref{fig:mcmc_fits_S0-102_like} show the constraints on the
spin parameters as a function of the observational duration for
S2-like and S0-102-like pulsars, respectively.  For such pulsars we
can see that the spin of the MBH can be constrained by $\DD a\sim0.1$,
$\DD i\sim 20-30^\circ$, $\DD \epsilon\sim 20-40^\circ$ within
$\sim4-8\,$yr.  The constraints on the spin can be correspondingly
tighter if the observations last longer than $\sim8-10\,$yr. If proper
motions with accuracies around $\sigma_{\rm p}=10\,\mu$as can also be
collected, the constraints could be tighter, by weakening the
degeneracies among the spin parameters due to the timing measurement
(See Section~\ref{subsubsec:degeneracy}). The constraints become $\DD
a\sim0.1$, $\DD i\sim 5-20^\circ$, $\DD \epsilon\sim 10-30^\circ$
within $\sim8$yr.

It is apparent that the most significant improvements on the
constraints occur after pericenter passages. 
In the most optimistic case, the observation starts near the pericenter
passages of the pulsar, in which case the spin can be constrained
within $\sim 2-4$yr for both the S2-like and S0-102-like pulsars. All
this is for a relatively low timing accuracy, i.e., $\sigma_T=5\,$ms.
In reality the timing accuracy could be accumulated to be $\sigma_T\la
0.1-1\,$ms. Thus the constraint of spin can be as fast as about
$2-3\,$yr in the most optimistic cases.

The constraints on spin parameters (especially the
  orientations) seem to become slightly weaker when the observation is
  before the second pericenter passage (see
  Figure~\ref{fig:mcmc_fits_S2_like} and
  Figure~\ref{fig:mcmc_fits_S0-102_like}). The reason is that we have
  fixed the number of observational samples at 120, instead of
  accumulating with time.  The spin-induced signals of the pulsar
  between the second apocenter and the next pericenter decrease as a
  function of time (see top left panel of
  Figure~\ref{fig:fchq_spin_Ea}).  This makes the overall spin-induced
  signals smaller if the number of data points is fixed.

\subsubsection{General cases}
\label{subsubsec:general}
%
\begin{figure*}
\center
\includegraphics[scale=0.8]{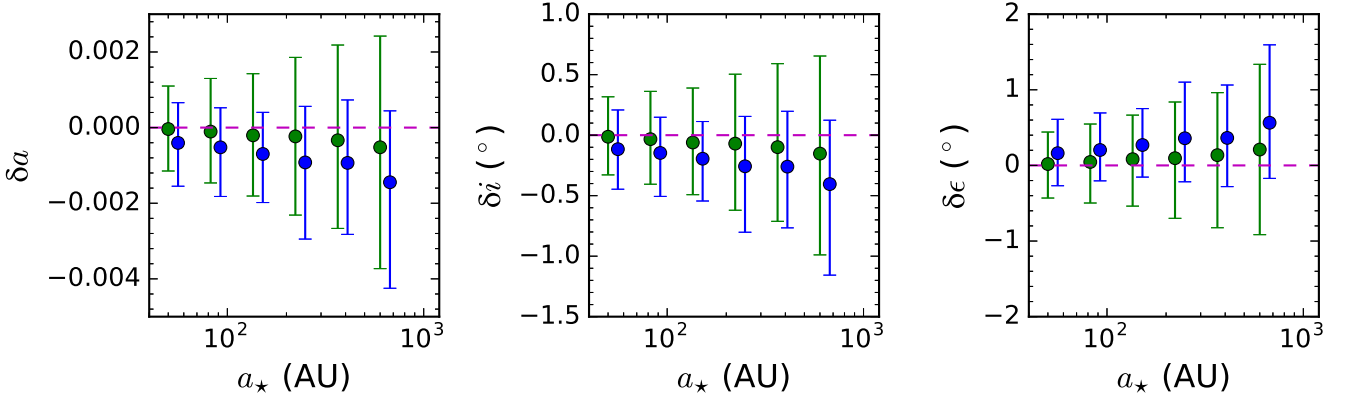}
\caption{Constraints on the spin and its orientation as a 
function of the orbital semimajor axis of the pulsar.  Here the pulsars are in
  orbits similar to Ea but the orbital semimajor axis varies from
  $50\AU$ to $600\AU$.  The assumed measurement errors are
  $\sigma_{\rm T}=5$ms and $\sigma_{\rm p}=10\mu$as. The green dots
  show the MCMC results when only the TOAs are used while the blue
  dots are results when both the TOAs and the proper motion are
  used. Small offsets in $x$-axis are used for the blue dots for
  viewing clarity.}
\label{fig:mcmc_fits_a}
\end{figure*}
\begin{figure*}
\center
\includegraphics[scale=0.8]{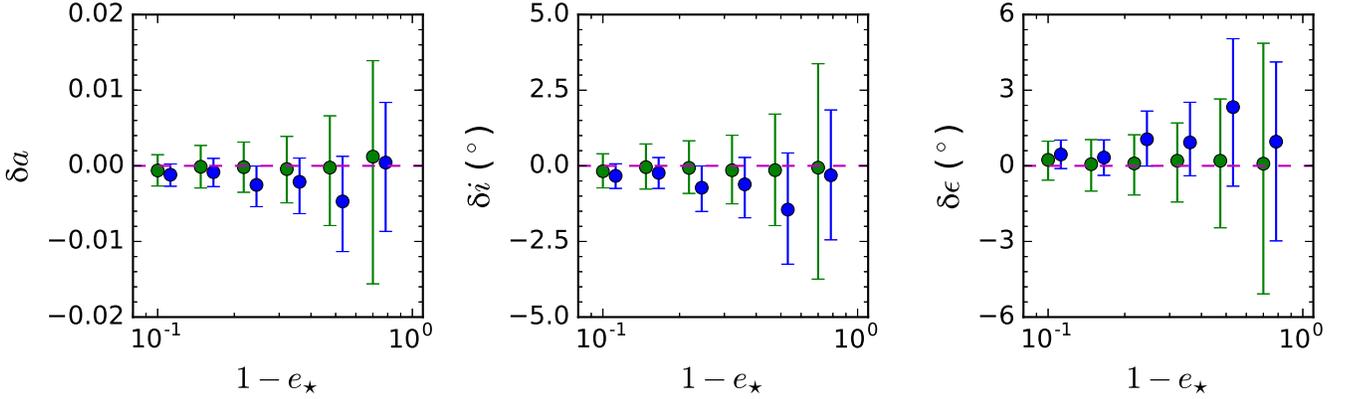}
\caption{Similar to Figure~\ref{fig:mcmc_fits_a}, but with the
  eccentricity of the pulsars varying from $0.3$ to $0.9$.}
\label{fig:mcmc_fits_e}
\end{figure*}

\begin{figure}
\center
\includegraphics[scale=0.5]{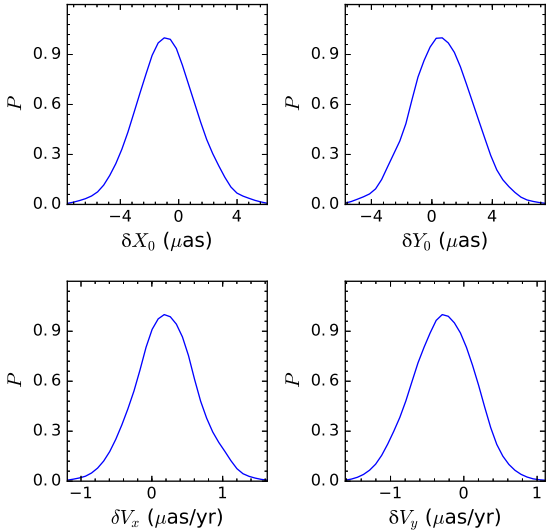}
\caption{Constraints on the position and proper motion of the Sgr~A* by observing both the timing and the apparent 
motion of the pulsar Ea in $8$yr. The assumed measurement errors are $\sigma_{\rm T}=5$ms and $\sigma_{\rm p}=10\mu$as.
}
\label{fig:cons_proper}
\end{figure}
\begin{figure}
\center
\includegraphics[scale=0.5]{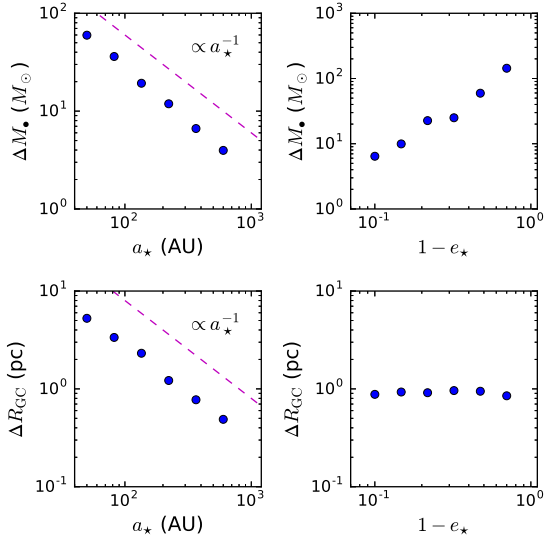}
\caption{Constraints on the MBH mass ($\DD\bh$, upper panels) and GC distance
  ($\DD R_{\rm GC}$, lower panels) when both the timing and the proper motions are
  used, as functions of $a_\star$ (left panels) or $e_\star$ (right panels) of the pulsar. The other initial 
  conditions of the pulsars are the same as Ea. The assumed measurement errors are $\sigma_{\rm
    T}=5\,$ms and $\sigma_{\rm p}=10\,\mu$as. The magenta lines in the top and bottom
  left panel show the reference scaling relations of $\propto a_\star^{-1}$. }
\label{fig:mcmc_fitsmr}
\end{figure}

If $\chi=\chi^Y_2$ denotes the value where $Y$ is at its $2\sigma$
boundary value obtained from the MCMC runs, according to
Equation~\eqref{eq:chqT} and~\eqref{eq:chqR}, we have, approximately,
\be\ba
&(\chi^{Y}_{2})^2\simeq 
\frac{N}{T_{\rm tot}}\int^{T_{\rm tot}}_0 \left\{\frac{\left[\phi(Y_0+\DD Y/2,t_{\rm arr})
-\phi(Y_0,t_{\rm arr})\right]^2}
{\nu_0^2\sigma_{\rm T}^2} \right.\\
+&\left. \frac{\left[\mathbf{R}(Y_0+\DD Y/2,t_{\rm arr})
-\mathbf{R}(Y_0,t_{\rm arr})\right]^2}
{\sigma_{\rm p}^2}\right\} dt_{\rm arr}\\
\simeq &
\frac{\DD Y^2}{4}\left[
\frac{1}{\nu_0^2\sigma_{\rm T}^2}\left(\overline{\frac{d \phi}{d Y}}\right)^2 
+\frac{1}{\sigma_{\rm p}^2}\left(\overline{\frac{d \mathbf{R}}{d Y}}\right)^2 
\right] N \left(\frac{T_{\rm tot}}{P}\right)^2\\
~\label{eq:chiy2}
\ea
\ee
Here $N$ is the total number of epochs with data, 
$P$ is the orbital period and 
$T_{\rm tot}$ is the duration of the observation. $\overline{\frac{d \phi}{d Y}}$
and $\overline{\frac{d \mathbf{R}}{d Y}}$ are the averaged 
derivatives per orbit defined similar to Equation~\ref{eq:avgdaY}.

Thus, if only the TOAs are used, we
have
\be
\DD Y\propto \left(\overline{\frac{d \phi}{d Y}}\right)^{-1}
\,\nu_0\,\sigma_{\rm T} \, N^{-1/2}
\left(\frac{T_{\rm tot}}{P}\right)^{-1}
\label{eq:DDY}
\ee
We can see that the constraint on $Y$ is approximately proportional to
$N^{-1/2}$ and $(T_{\rm tot}/P)^{-1}$, 
i.e., the constraints will be
improved if more epochs are observed and if the observations last for
a longer time.  Also, the constraints can be improved if the intrinsic
spinning frequency of the pulsar $\nu_0$ are higher. Note that $N$
should be much larger than the number of free parameters in MCMC simulations to
ensure a meaningful fit.  

We also explore the constraints on the spin parameters for pulsars
with different $a_\star$, $e_\star$. Figure~\ref{fig:mcmc_fits_a}
shows the constraints on spin from pulsars similar to Ea, but with
various $a_\star$, assuming $\sigma_{\rm T}=5\,$ms.
We can see that $\DD a=0.002-0.006$, $\DD i=0.6-2.0^\circ$,
$\DD\epsilon=0.9-3.0^\circ$ (or $\DD a=0.002-0.005$, $\DD
i=0.6-1.0^\circ$, $\DD\epsilon=0.9-2.0^\circ$) if only the TOAs are
used (or if both the TOAs and the apparent positions are used), when
$a_\star$ varies from $50\,\AU$ to $600\,\AU$.  We find that $\DD a$
is approximately proportional to $a_\star^{1/2}$.  It can be
understood, as according to Equation~\eqref{eq:DDY}, approximately 
$\DD a\propto (d\phi/da)^{-1}
\propto (\delta_a t_{\rm arr})^{-1} \propto a_\star^{1/2}$ 
(See also Section~\ref{subsubsec:pulsar_df_orbits}).

Figure~\ref{fig:mcmc_fits_e} shows the constraints on spin for pulsars
similar to Ea, but having various $e_\star$. We can see that $\DD
a=0.03-0.004$, $\DD i=7.0-1.0^\circ$, $\DD\epsilon=10.0-2.0^\circ$.
(or $\DD a=0.02-0.003$, $\DD i=4.0-0.8^\circ$,
$\DD\epsilon=7.0-1.0^\circ$) if only the timing signals are used (or
if both the timing and the position observations are used), when
$e_\star$ varies from $0.3$ to $0.9$.

These results suggest that pulsars with distance $<600\AU$ and
$e_\star\ga0.3$ can set very tight constraint on the spin in three
orbits.  For these pulsars, note that the constraints on spin by
additionally using the astrometric data are only modestly
improved. This is because the constraints on spin parameters are
dominated by the parts from the TOA, i.e.,
$\chi_{\rm T}\gg\chi_{\rm P}$.

\subsection{Constraints on the mass, distance and the proper motion of the MBH}
~\label{subsec:MCMC_results_mr}

Using both the astrometric measurements and the TOAs of pulsars in the
MCMC simulations can provide precise estimate of the mass, distance and the proper
motion of the MBH.  The constraints on proper motion of the MBH for Ea
are shown in Figure~\ref{fig:cons_proper}.  We can see that the
position of the MBH can be constrained to accuracies of $\DD
X_0=8.1\mu$as, $\DD Y_0=8.0\mu$as, $\DD V_x=1.7\mu$as$/$yr, $\DD
V_y=1.6\mu$as$/$yr. The constraints on the mass and the GC distance
are $\DD M=10.6\msun$ and $\DD R_{\rm
  GC}=1.2$pc. Figure~\ref{fig:mcmc_fitsmr} shows constraints on the
mass and the GC distance for pulsars in different orbital semimajor
axes and eccentricities.  From Figure~\ref{fig:mcmc_fitsmr} we can see
that the mass can be constrained by a factor of $\sim
10^{-5}-10^{-6}$, i.e., $\DD \bh=1\sim10^2\msun$. The GC distance is
constrained to an accuracy of $\sim 10^{-4}-10^{-3}$, i.e., $\DD
R_{\rm GC}=0.5\sim 5$pc.  Interestingly, we find that the constraints
on MBH mass and GC distance are more accurate for pulsars at larger
distances from the MBH. 
The main reason is that relative position (or timing) error 
is inversely proportional to the semimajor axis of the star: 
Approximately, $\phi\propto \Delta_{\rm R} \propto a_\star$ and $|\mathbf{R}|\propto 
a_\star$, thus according to Equation~\ref{eq:chiy2} we have $\DD \bh \propto 
a_\star^{-1}$, $\DD R_{\rm GC}\propto a_\star^{-1}$.
These scaling relations are well consistent with those obtained by the 
MCMC simulations (See Figure~\ref{fig:mcmc_fitsmr}).

Pulsars with high orbital eccentricities can
help to put tighter constraints on the MBH mass, however, that
does not appear to help in constraining the GC distance.

\subsection{The effects of the pulsar's mass}
\label{subsec:effect_pulsar_mass}
In our Kerr metric framework, the pulsar is a test particle and thus
its mass is ignored. The difference in the pulsar timing due to
this approximation, if there are any, should be of the order of the mass ratio, 
i.e., $10^{-7}-10^{-6}$ as the pulsar's mass is $m\sim1.4\msun$. However, 
considering that the timing accuracy of the pulsar is quite high, 
such differences could be detectable. Nevertheless, 
we find that the simulation results and conclusions in this work are
only slightly affected by the pulsar's mass. 
The details of the estimations and the discussions are as follows.

In Newtonian physics, the orbital period of the pulsar is determined by 
the total mass of the binary. Thus, we expect that the measured mass of the MBH 
from our MCMC simulations should be effectively the sum of the 
true mass of the MBH and the pulsar. As the accuracy of MBH's mass obtained 
by GC pulsars is $\simeq4-10^2\msun$ (See top left panel of Figure~\ref{fig:mcmc_fitsmr}), 
the bias of the estimated MBH's mass should not be detectable,
unless the pulsar is relatively far away ($>1000\AU$).

Both the Roemer delay and the Einstein delay are affected by the mass of the pulsar. 
Note that the Shapiro delay does not, as it depends only on the mass of 
the MBH. Approximately, we have $\Delta_R\propto 1-m/\bh$ and 
$\tilde\gamma\propto 1-\frac{1}{2}m/\bh$~\citep{Damour86}, where $m$ is the mass
of the pulsar. For GC pulsars, we have $\Delta_R=10^5-10^6$\,s and $\tilde\gamma=10^4-10^5$\,s
(See Figure~\ref{fig:ftiming}), thus, the difference due to ignoring the pulsar 
mass should be $0.01-1$\,s and $1-100$\,ms for the Roemer delay and the Einstein delay, 
respectively. These differences may lead to biases on the parameters of the pulsar-MBH 
binary estimated by the MCMC simulations. Considering that the spin-induced effects are in orders 
of $10-100$\,s (See Figure~\ref{fig:fchq_spin_Ea} or~\ref{fig:fchq_spin_Eb}, 
   top left panel), we expect that the spin parameters, if they are affected,
should not be significantly biased from their true values. 

The relativistic effects in the timing of pulsars are originated from the 
orbital precessions, which are also affected by the pulsar's mass. For example, 
The Schwarszchild orbital precession is proportional to $1+m/\bh$ and 
spin-induced orbital precessions are proportional to $1+\frac{7}{4}m/\bh$~\citep{Wex99}.
Thus, the bias of these effects should be of the order of the mass ratio, 
i.e., $10^{-6}-10^{-7}$, which can be ignored.

We note that the effect of the pulsar's mass can be included by 
introducing a corresponding perturbative term into the Hamiltonian. However, 
deriving the explicit form of this term and such extensions of the current numerical method
are beyond the scope of this work. We defer them to future studies.

\section{Discussion}
Our results suggest that the spinning magnitude of the MBH can be constrained down to $\sim10^{-2}$
within a decade even if the timing accuracies are relatively low, i.e., $\sigma_{\rm T}\sim5$ms.
This suggests that SKA1-MID, not necessarily the final stage of SKA2, can already probe
the spinning nature of the GC MBH, if any pulsar within $\la1000\AU$ can be found. 
Indeed, it is suggested that SKA1-MID is probably able to reveal the hidden pulsars 
at as low as $2.4$GHz with spin period $\sim 0.5$s in this region~\citep{Eatough15}.

Our simulations have the advantage that all quantities are obtained under the Kerr metric. 
The signals could be more accurate than those of the previous studies that are based on post-Newtonian 
approximation methods. The main disadvantages of our method are that the gravitational wave decay is not included,
and that the torque effects are not easy to be discussed separately. 
As a consequence, alternative-gravity theories are hard to discuss in a theory-independent way. 
Anyway, the deviations of the GR can still be detected by comparing our model predictions with 
the observables, if the full GR model could not fit the observations well, i.e., 
if the MCMC parametric fits leave significant residuals. 

Note that some GR effects unique to pulsar timing are not covered in 
this work. For example, the shift of the time of emission of the pulse centroid due to the 
spinning precession of the pulsar, the distortion of the pulse profile~\citep[e.g.,][]{Rafikov06}, 
or the high order pulses due to extremely strong gravitational bending~\citep{Wang09}. 
These effects can be straightforwardly included in our framework as the geodesic equation and the 
light trajectories have been solved explicitly in our method. However we notice that some of these effects are significant 
only if the pulsars are in edge-on orbit~\citep[e.g.,][]{Rafikov06}, thus they have negligible effects
for results shown in this study. 

The pulsars could be perturbed by other surrounding gravitational sources, e.g., the pulsars or other
stellar remnants. The effects of the background perturbation are expected to be important outside 
$\ga100-400\AU$~\citep{Merritt10,Zhang16}. Due to the different nature of these background perturbations, 
they are expected to be separable from the GR effects~\citep{Angelil14,Zhang16}. 

\section{conclusions}
It is believed that pulsars rotating closely around the GC MBH are superb tools in probing the GR and 
the gravity theories. Based on a relativistic framework developed in our 
previous work, here we study both the TOA and the apparent motion of these pulsars and the corresponding 
spin-induced effects. We take the pulsar as a test particle and solve explicitly the geodesic
equations of the pulsar's motion and its pulse trajectories to the observer in the Kerr metric. 
By performing a number of MCMC simulations, we investigate the constraints on the 
spin and other properties of the GC MBH achievable by monitoring surrounding pulsars.

We find that the full GR treatment is necessary in describing accurately the 
timing signals. If approximate models are used, that assume that the orbital precession increases 
linearly with time, the predicted TOA difference due to spin effects can deviate from the results of 
our relativistic simulations up to $\sim10\,$s, which would be quite apparent for timing observations 
performed by future facilities, e.g., the SKA.

We find that the spin-induced TOA differences can mount up to $\sim140$\,s in $\sim8\,$yr
for a pulsar with orbital period of $\sim2.6$\,yr. Even for S2-like or S0-102-like 
pulsars, the spin-induced TOA differences can be up to $\sim80$\,s ($\sim20$\,s) after $40$\,yr ($30$\,yr) 
of observation. The signal is orders of magnitude larger than the timing accuracies expected in the future 
($\la1-10$ms), thus it should be possible to set tight constraints on the spinning of the MBH. 

We perform a number of MCMC simulations to study the constraints on the spinning of the MBH.
We find that strong near-degeneracies among the spin parameters could appear, if only the timing of pulses 
are used. Such near degeneracies can be weakened if the pulsar proper motion is measured  with considerable accuracies 
along with the timing, or if the pulsar is close enough to the MBH such that 
the orbital precession or the quadrupole-moment effects are significant.

Although near degeneracies exist if only the timing of pulsars are used, the constraints of the spinning 
parameters are still very tight. By monitoring a normal pulsar with orbital period of 
$\sim2.6$yr and eccentricity of $0.3-0.9$, and assuming the timing accuracy of $1-5$ms, 
we find that within $\sim 8$yr the magnitude, the line of sight inclination and the position 
angle of the MBH spin can be constrained with $2\sigma$ error given by $10^{-3}-10^{-2}$ and 
$10^{-1}-5^\circ$, $10^{-1}-10^\circ$, respectively. 

Even for pulsars in orbits similar to the currently detected star S2/S0-2 or S0-102 and providing that the 
timing accuracy is $\sim5\,$ms, we find that the spinning of the MBH can still be constrained within $4-8\,$yr.
The most significant constraints of the spin parameters are provided near pericenter passage. Thus, in the optimistic case 
that the timing observations start near the pericenter passages of pulsars, the spinning of the MBH 
can be constrained within $2-4\,$yr.

If the proper motion of the pulsars with accuracy of $10\mu$as can also be collected along with the 
timing measurement, then the position, velocity, mass and the distance of the MBH can be constrained 
about $\sim10\mu$as, $\sim10\mu$as$/$yr, $\sim 1\msun$ and $\sim1$pc, respectively. 

\acknowledgements
\noindent We are very grateful to the anonymous referee for all the suggestions have 
given to us, which have improved this paper significantly. We thank Zhu Weishan for providing the computing 
resource to the TianHe II National Supercomputer Center in Guangzhou, on which part of the simulations are performed. 
This work was supported in part by the National Natural Science Foundation of China under grant 
No. 11603083, 11673077. This work was also supported in part by ``the Fundamental Research 
Funds for the Central Universities'' grant No. 161GPY51, the Key Project of the 
National Natural Science Foundation of China under grant No. 11733010. Part of the numerical work 
was performed in the computing cluster in School of Physics and Astronomy, Sun Yat-Sen University.

\end{document}